\documentclass[%
reprint,
superscriptaddress,
 prl,
 amsmath,
 amssymb,
 aps,
 longbibliography
]{revtex4-2}

\usepackage{cancel}
\usepackage{graphicx}
\usepackage{dcolumn}
\usepackage{bm}
\usepackage[usenames,dvipsnames]{color}
\usepackage{xcolor}
\usepackage{soul}
\usepackage{enumitem}
\usepackage{xspace}
\usepackage{url}
\usepackage{booktabs}
\usepackage{lipsum}
\usepackage{siunitx} 
\usepackage[ISO]{diffcoeff}
\usepackage{subcaption}
\usepackage{ragged2e}
\usepackage{mathrsfs}
\usepackage{hyperref}
\usepackage[capitalise]{cleveref}
\usepackage[normalem]{ulem}
\usepackage{amsmath}
\usepackage{esvect} 
\usepackage{titlesec}
\usepackage{titletoc}
\usepackage{chngcntr}
\usepackage{fontawesome5}
\usepackage{mleftright}
\usepackage[nodayofweek]{datetime}
\usepackage{comment}

\usepackage{gensymb}

\DeclareSIUnit\year{yr} 
\DeclareCaptionJustification{justified}{\justifying}
\definecolor{firebrick}{HTML}{B22222}
\definecolor{green2}{RGB}{89, 122, 0}
\definecolor{green3}{RGB}{13, 66, 14}
\definecolor{green4}{RGB}{38, 74, 0}

\hypersetup{
    colorlinks=true,       
    linkcolor=firebrick,          
    citecolor=firebrick,        
    filecolor=firebrick,      
    urlcolor=firebrick          
}

\definecolor{orcid-green}{RGB} {166, 206, 57}
\newcommand{\MYhref}[3][blue]{\href{#2}{\color{#1}{#3}}}%

\titleclass{\mysection}{straight}[\section]
\titleformat{\mysection}[runin]
  {\itshape}{\thesection}{}{}[.---]
\titlespacing{\mysection}{1em}{1em}{0em}

\DeclareSIUnit\clight{c}

\begin{document}

\title{Detection of the Earth Tides by Diamagnetic Levitation}

\author{Tim M. Fuchs\,\MYhref[orcid-green]{}{\faOrcid}}
\affiliation{School of Physics and Astronomy, University of Southampton,
SO17 1BJ, Southampton, UK}

\author{Elliot Simcox\,\MYhref[orcid-green]{}{\faOrcid}}
\affiliation{School of Physics and Astronomy, University of Southampton,
SO17 1BJ, Southampton, UK}

\author{Michael Hawker\,\MYhref[orcid-green]{}{\faOrcid}}
\affiliation{School of Physics and Astronomy, University of Southampton,
SO17 1BJ, Southampton, UK}

\author{Francis J. Headley\,\MYhref[orcid-green]{https://orcid.org/0009-0000-7585-4957}{\faOrcid}}
\affiliation{Institut für Theoretische Physik, Eberhard-Karls-Universität Tübingen, 72076 Tübingen, Germany}

\author{Hendrik Ulbricht\,\MYhref[orcid-green]{https://orcid.org/0000-0003-0356-0065}{\faOrcid}}
\email{Correspondence to: H.Ulbricht@soton.ac.uk}
\affiliation{School of Physics and Astronomy, University of Southampton,
SO17 1BJ, Southampton, UK}

\begin{abstract}
\noindent
The detection of mass distributions and mass transport via gravity mapping is a key geophysical tool for understanding Earth’s structure and dynamics. Changes in mass distribution, driven by natural processes and human activity (e.g., extraction of oil, gas, and minerals), contribute to observable phenomena such as sea-level rise (\SI{3}{mm\,yr^{-1}}), increased flooding, landslides, and ice mass loss (hundreds of \SI{}{Gt\,yr^{-1}}). These processes generate gravity variations detectable by gravimeters and gradiometers on ground and in space. Current instruments achieve sensitivities of ~10-100\,\SI{}{\micro Gal\,Hz^{-1/2}} and enable applications including hydrocarbon exploration, volcanic monitoring, and subsurface detection. They also measure Earth tides (100-300\,\SI{}{\micro Gal} amplitude), requiring long-term stability over days. However, existing systems are limited by size (\textgreater 8 kg) and cost (\textgreater10$^5$ USD), restricting widespread deployment.
Here we demonstrate a levitated optomechanical sensor (LOMS) with a demonstrated sensitivity of \SI{18}{\micro Gal}, a large dynamical range, and an integration time of \SI{6}{s}, with an expected sensitivity of $\leq$ \SI{200}{n Gal\,Hz^{-1/2}} in a volume of only a few \SI{}{cm^3}. We resolve Earth tide signals, demonstrating stability comparable to state-of-the-art instruments. Unlike conventional accelerometers (\SI{}{\micro g} sensitivity, low stability), our device operates as a true gravimeter. Its compact size and low projected cost enable scalable deployment, including drone-based surveys (10–100 \SI{}{m} altitude), distributed sensor networks, and multi-pixel gravity imaging arrays. This platform enables high-resolution, cost-effective gravity mapping with potential for large-scale geophysical monitoring.
\end{abstract}
\maketitle

\paragraph{\textbf{Introduction ---}}
\label{sec:intro}
    Sensitive gravimeters are of significant economical and scientific importance, as the mapping of the local gravitational field has been used extensively for the detection of natural underground features such as reservoirs~\cite{Tucci_1983}, oil and natural gas deposits~\cite{newaz_2023} and geological features~\cite{arisona_2023} such as mineral deposits~\cite{mining5010003}. Man-made features have also been detected, both for the field of archaeology~\cite{Panisova_2009} and scanning underground facilities~\cite{Romaides_2001,stray_lamb}. Such gravitational maps are commonly created terrestrially and locally, either by land based sensors~\cite{Murzabekov_2025} or low flying aircraft~\cite{forsberg2010airborne,Scheinert_2016}, with rarer examples created by ship based sensors~\cite{bidel_zahzam}. However, satellites such as GRACE~\cite{Tapley_2004}, GOCE~\cite{pail_2011} and LISA-PF~\cite{Armano_2016} can create and update gravitational maps of the entire planet. As well as the detection and analysis of features in these maps, gravitational maps can be used for Gravity-Aided Inertial Navigation Systems (GINS) by matching the detected local gravity to known gravitational maps, which are more robust than pure GPS based navigational systems~\cite{AHMADIAN2026120161}.
    The change in local gravity over time is of interest for a variety of reasons such as monitoring seismic activity in sensitive locations such as observatories~\cite{VERMA2023100165}, monitoring the magma in volcanoes to give advance warning of eruptions~\cite{CARBONE2017146,Rymer_2000}, and monitoring the status of underground reservoirs~\cite{Nishijima_2016,Gasperikova_2008}.\\
    Most gravimeter use cases in sensing and navigation require sensors to be robust and economical, while still retaining sensitivity. Furthermore, long term sensor stability limits time-consuming calibration, increasing sensor duty cycle in the field.\\
    The current world leading gravimeters use superconducting levitation to measure spring-like forces on a levitated mass~\cite{Warburton_2010} and can achieve precisions of \SI{1}{nGal}~\cite{Riccardi_2011} with drifts as low as \SI{1.6}{\micro Gal/year}~\cite{CROSSLEY2004325}. Sensitive gravimeters systems are large, expensive, and power intensive, limiting their adoption. In recent years, micro-electromechanical-system (MEMS) gravimeters have been developed which can be more compact and efficient than traditional superconducting gravimeters. They have not yet reached competitive sensitivities or stability characteristics but are at the level demanded by some industrial applications, currently reaching resolutions below \SI{10}{\micro Gal}~\cite{mustafazade_2020,Middlemiss2016,wu_2025,gao_wu}. While MEMS devices are more power efficient than their superconducting competitors, they still require total powers in excess of \SI{500}{W} and need to be kept at a temperature stable to within \SI{\pm 0.1}{mK}~\cite{carbone2020newton}. 
    Atom interferometer gravimeters have costs and sizes comparable to superconducting gravimeters whilst having demonstrated sensitivities only marginally better than MEMS devices~\cite{portableAI}. However, they offer comparably smaller drift and, like free falling masses, can give absolute gravity measurements~\cite{AIreview2024}, as opposed to the relative measurements offered by many other types of gravimeter. These free falling mass gravimeters, such as corner-cube absolute gravimeters, can accurately measure absolute gravity, and concurrently, vertical gravity gradients. However, they need fine alignment during setup, are especially sensitive to vibrations, and are relatively bulky~\cite{corner_cube}. \\
    In this paper we present a gravimeter utilizing levitated physics, with low power and size requirements by low-noise passive trapping of diamagnetic material above a array of permanent magnets.
    Levitated systems offer large acceleration sensitivity and excellent decoupling from external sources of noise~\cite{timberlake_2019,vinante_2020, fuchs2024}, where their use has been proposed for fundamental physics experiments on precision measurements of Newtons constant~\cite{headley2025}, gravitationally mediated entanglement~\cite{carlesso_2017,weiss_2021}, and forces beyond the standard model of particle physics~\cite{Amaral5F}. Here we demonstrate levitated mechanical sensors for gravimetry in relevant real-world applications.\\
    The weak confinement of levitated systems directly translates to low resonance frequencies and large force sensitivities in the sub-resonant regime. As seismic activity is predominantly low-frequency, levitated systems are well suited to monitoring these signals.
    Here we show that diamagnetic levitation, compared to other levitation schemes, benefits from being a passive trapping scheme, removing much of the noise and drift introduced by common active levitation schemes. Furthermore, diamagnetic schemes allow for the levitation of significantly heavier test masses when compared to the most widely used form of levitation, optical levitation. The added mass in turn greatly lowers the complexity of the detecting set-up, significantly reducing the cost of the sensor presented in this paper. 
    Tidal forces are used as a low frequency benchmark for the force sensitivity of our levitated optomechanical sensor (LOMS), containing lunar and solar components (Period of \SI{12.42}{h} and \SI{24}{h}, respectively) in the order of \SI{E-4}{Gal} and \SI{5E-5}{Gal}, respectively.\\

\begin{figure}[!t]
        \centering    
        \includegraphics[width=\columnwidth]{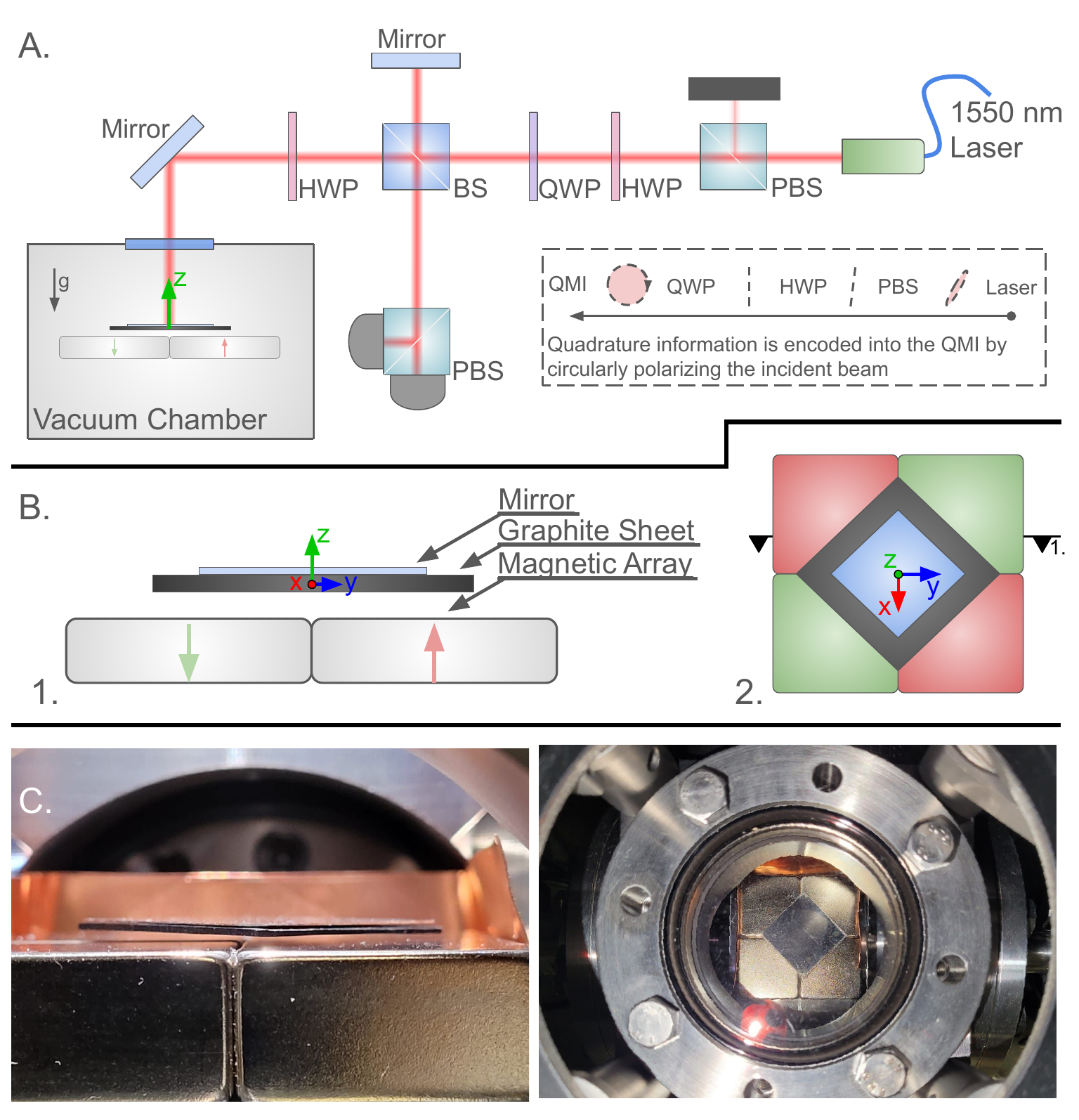}
        \caption{\textbf{Schematic of the levitated optomechanical sensor --- A:}~Schematic of the experimental set-up, showing the Quadrature Michelson Interferometer (QMI) readout and the diamagnetically levitated sheet of graphite with mirror on top. The resonator is placed inside of a vacuum chamber to isolate it from air currents. The entire system is placed on an active vibration isolation stage to further limit seismic noise. The inset shows how the polarization of the beam is changed as part of the QMI. \textbf{B:}~Schematic of the magneto-mechanical resonator. A sheet of graphite (HOPG) is levitated above a magnetic Halbach array through the diamagnetic repulsion of the graphite. A mirror is placed on top of the graphite sheet to increase reflectivity in the measurement arm. The direction of the magnetic moment is shown in green (down) and red (up). \textbf{C:}~Images of the resonator levitated within the set-up, from left to right: Side-view of the mirror placed on HOPG, which is levitated above the magnetic array; top-view of the sensor inside of the vacuum chamber.}
        \label{fig:setup}
    \end{figure}

\paragraph{\textbf{Method ---}}
\label{sec:theory}
    Our gravimeter is formed by a sheet of highly oriented pyrolytic graphite (HOPG) with a mirror (total mass, m = \SI{1.19e-3}{kg}) diamagnetically levitated above an array of rare-earth magnets in a Halbach configuration, shown in Fig.~\ref{fig:setup}B. This levitation creates a soft spring with a resonant frequency of \SI{0.81}{Hz}, with spring constant $k_z = m\omega_{n,z}^2 = \SI{31}{mN/m}$ ($k_z \approx 51\,k_{x,y}$ according to our model, see the next section). The natural frequency also forms the effective duty cycle of the sensor. Below this frequency, the spring constant forms the flat force-displacement transfer function.
    By measuring the displacement $q\in(x,y,z)$ of the graphite, the force experienced by the graphite sheet can be determined using Hooke's law, $F_q = k_q~q$, combining this with Newton's second law, we find the acceleration experienced by the sensor as    
    \begin{equation}
        a_q = (2\pi f_{n,q})^2~q,
        \label{eq:displacementconversion}
    \end{equation}    
    where $f_{n,q}$ is again the natural frequency of the levitation along the axis $q$. In the case of tidal effects, the DC position of the test mass is shifted by the gravitational pull of celestial objects. These tidal effects are typically much lower frequency than our duty cycle. To accurately measure the acceleration experienced by our sensor, we employ optical interferometric readout of the displacement. The interferometric readout is most sensitive along the $z$ axis of the system, but the shape of the trapping potential and small angular effects lead to the detection of in-plane accelerations as well.
    
   The quadrature Michelson interferometer (QMI) is schematically depicted in Fig.~\ref{fig:setup}A. We use circularly polarized light, which is sent into the interferometer, passing through a non-polarizing beam-splitter and reflected off of a stable reference mirror and a mirror placed on top of our graphite sheet. The reflected light is recombined, and then decomposed into two orthogonal components by a polarizing beam splitter, each component incident onto a separate photodiode detector, to create a quadrature signal. Since the orthogonal components in fully circularly polarized light have a~\SI{90}{\degree} phase shift with respect to each other, this provides an optical quadrature displacement measurement, extending the working range of the displacement measuring interferometer beyond the half-wavelength fringe of a conventional Michelson interferometer, to half the coherence length of the laser.
    We introduce a secondary half-wave plate in the measurement arm to correct the slight polarization change induced by the window in this arm of the interferometer.
    Using tip-tilt stages, the fringe visibility of the interference signals are maximized and the quarter-wave plate in the measurement arm is used to ensure the displacement signal is circular in quadrature after the initial beam is verified to be fully circular going in, for maximal sensitivity over the displacement range.
    
    In order to convert the voltages measured by our interferometer to acceleration, we use equation~\ref{eq:displacementconversion}. The conversion from voltages to displacement is discussed in Supp.~A..
    
    We consider two main noise sources. The first being the resonators thermal force noise, expressed in acceleration noise, 
    \begin{equation}
        S_{a,th}^{1/2} = \sqrt{4k_B T\frac{k_q}{m^2\omega_{n,q}Q_q}},
        \label{eq:thermalnoise}
    \end{equation}
    where $k_B$ is the Boltzmann constant, T the temperature of the resonator, $\omega_{n,q}$ and $Q_q$ are, respectively, the natural angular frequency and mechanical quality factor of motion along the $q$ axis of the system, and m the mass of the graphite under test. The other source of noise considered is the voltage noise present in the readout,
     \begin{equation}
        S_{a, det}^{1/2} = \frac{k_q\lambda \pi S_V^{1/2}}{m V_{max}},
        \label{eq:detectionnoise}
    \end{equation}
    where $\lambda/2$ is a single period of displacement in the interferometer, with $\lambda$ the wavelength of the laser, $ V_{max}$ is the depth of the periodic voltage response corresponding to this displacement and $S_V^{1/2}$ is the single-sided voltage noise spectral density in units \si{V/\sqrt{Hz}}. Since we are studying seismic signals in this paper, we do not categorize them as noise, but rather as signal, otherwise this would be the dominant contribution to our noise. The long term drifts of the setup are characterized in the next section.\\

\paragraph{\textbf{Analysis and Results ---}}
\label{sec:anal}
     
\begin{figure*}[!t]
        \centering    \includegraphics[width=2\columnwidth]{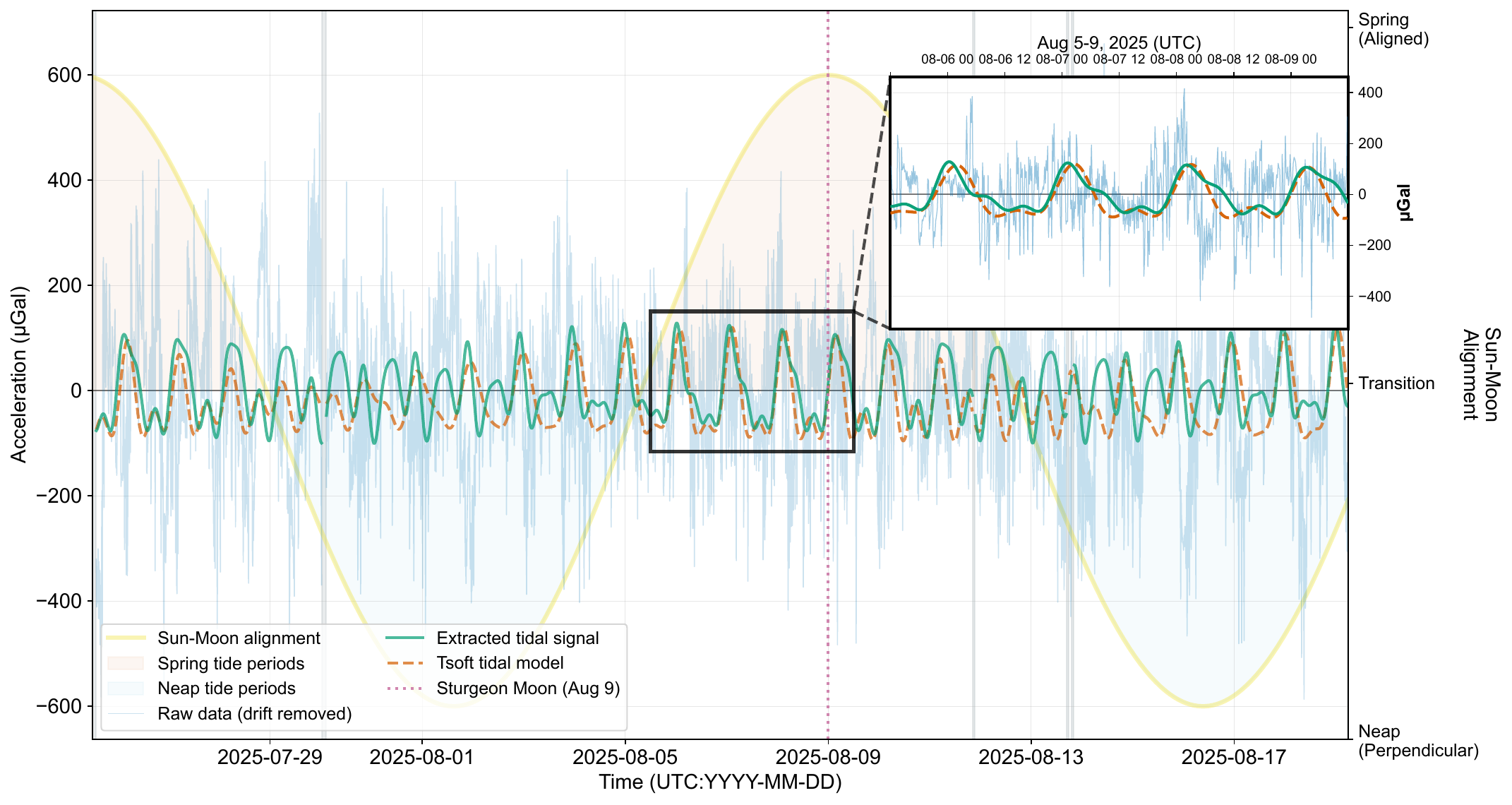}
        \caption{\textbf{Tidal signal detection of spring-neap cycle (Southampton, UK) ---} One month time trace of acceleration experienced by the Levitated Optical Mechanical Sensor, in units of \si{\micro Gal}, including the gravitational acceleration predicted by the Tsoft Model. The blue and red shaded regions indicate the lunar phase, which represents the alignment between the solar and lunar components of the tidal acceleration. The inset shows a zoom for the period between the 5th and 9th of August, during which the Sun and Moon where aligned along the same axis. The shoulder not predicted by the Tsoft Model is attributed to the 3D behavior of our sensor. The raw acceleration data before any filtering is applied is shown in light blue behind the green filtered signal. Grey shaded regions are used to mark where anomalous spikes in the raw data have been masked.}
        \label{fig:TimeTrace}
    \end{figure*}

    Fig.~\ref{fig:TimeTrace} shows a month long time trace of the acceleration measured by our device. We compared our data to the Tsoft-model~\cite{VANCAMPTSoft}, which is the time-series analysis tool for gravity data, which uses theoretical Earth-tide models in a geocentric reference frame. Tsoft is commonly used to analyze gravimeter data~\cite{Middlemiss2016}. The tidal signal from our LOMS device is extracted from the raw acceleration data by means of spectral filtering, which we discuss in detail in Supp.~B. 
    We find good agreement between Tsoft modeling and our experimental data. However, our device is sensitive to the full gravity vector and can differentiate between horizontal and vertical tidal effects, although in the current iteration, our detection can not decouple those effects. Depending on the relative constellation of the Earth, Moon and Sun, we observe deviations between Tsoft and our sensor signal. Tsoft only provides the gravitational component normal to the Earth's surface. Since our sensor can distinguish between the lunar and solar tides, this leads to differences between the acceleration measured by our sensor and the acceleration predicted by the Tsoft model when the sun and the moon are not aligned. To illustrate this, we superimpose the lunar phase on top of our time trace in Fig.~\ref{fig:TimeTrace}. Red-shaded regions correspond to the moon and the sun being more aligned along the same vector, whereas blue-shaded regions show the moon and the sun being less aligned, being perpendicular at the dip. As can be seen by comparing the red dashed tidal model to our green extracted data, the agreement between the model and the measured acceleration is significantly higher within the red-shaded regions. In this paper we limit ourself to discussing the normal-component, leaving the analysis of the full-vector sensitivity for future work. The inset of Fig.~\ref{fig:TimeTrace} shows a zoom-in for good alignment between the two tidal components. Our LOMS device detects an additional shoulder on the peak predicted by the Tsoft model, which we attribute to the horizontal motion of our sensor. 3D modeling will be used in a future study to resolve the detailed signal features of our sensor, while the current work focuses on the sensitivity and stability of the LOMS device.\\  

    \begin{figure}[t!]
        \centering    \includegraphics[width=\columnwidth]{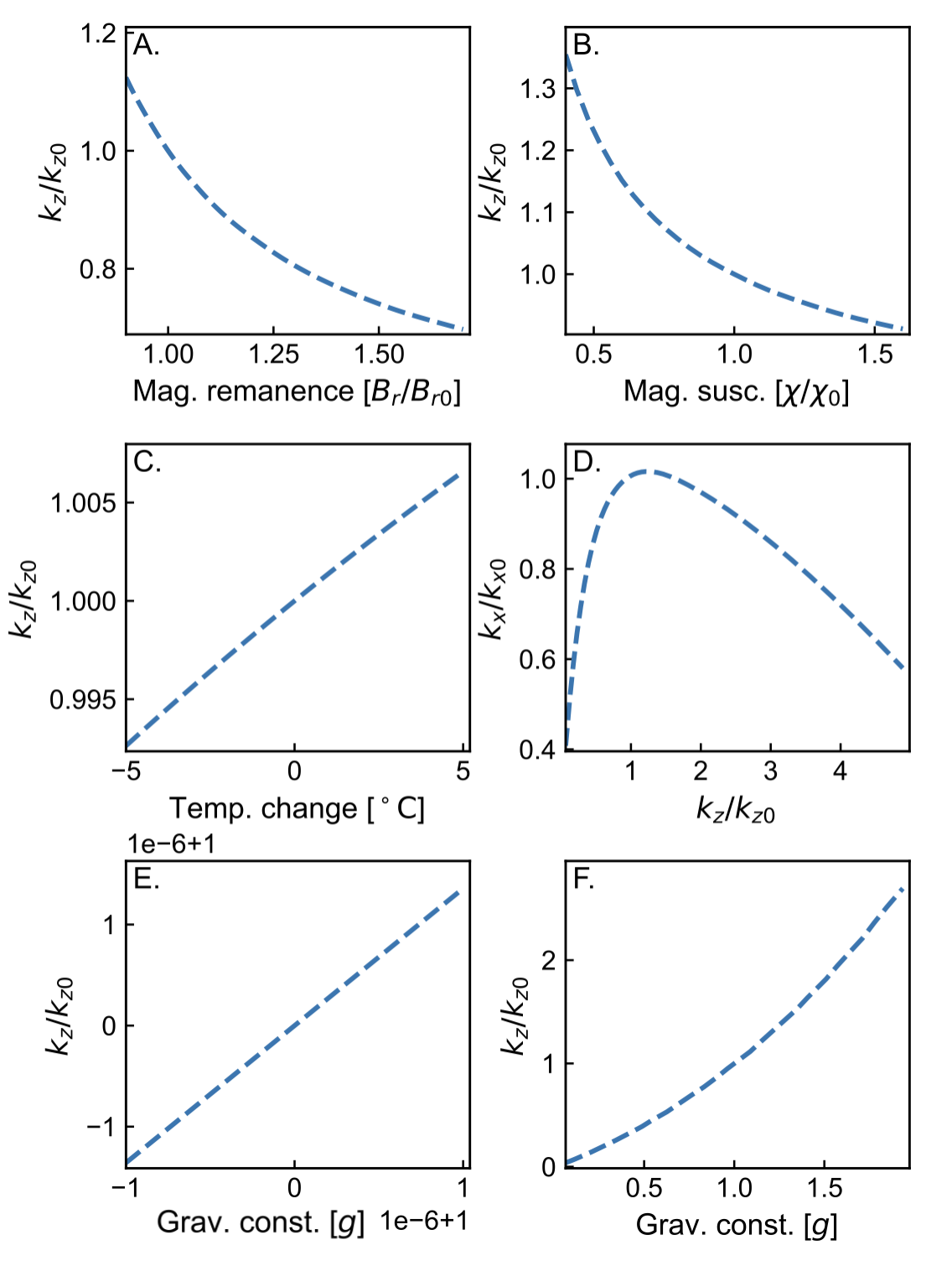}
        \caption{\textbf{The fractional spring constant as a function of different fractional experimental parameters ---} \textbf{A.} and \textbf{B.} demonstrate the decrease in the $z$-mode spring constant as the magnetic remanence and magnetic susceptibility increase, due to the increase in the levitation height. \textbf{C.} demonstrates the change in the spring constant as a function of the temperature, as both the magnetic remanence and susceptibility decrease linearly as the temperature increases. \textbf{D.} shows the parametric dependence of the $x$-mode spring constant as a function of $k_z$, where we have only varied the mass (i.e. a direct modulation of the levitation height). In plots \textbf{E.} and \textbf{F.} we see the increase in the spring constant as we increase $g$, with an almost linear relation for small perturbations of $g$. A table of the baseline parameters and exact scaling of the fractional parameters is given in the supplemental material.}
        \label{fig:springstiffness}
   \end{figure}

\paragraph{\textit{Numerical Model ---}}
    Since the sub-resonant sensitivity of our detector is given by the spring constant of the resonator and our minimum displacement sensitivity, we made a numerical model to see how the spring constant is affected by different experimental parameters. This model can be used to optimize the sensitivity of our sensor and explore what physical noise sources most strongly affect our long and short term stability.    
       
    The dependence and stability of our sensor on the parameters of the system were calculated numerically (see Supp.~C.). The Halbach array was modeled as a checkerboard of cuboidal magnets \cite{camacho_2013} and the components of the magnetic field were integrated over the surface area of the graphite sheet,    
    \begin{equation}
        U_\text{mag}(x,y,z)=-\dfrac{\delta}{2\mu_0}\int dA\left[\chi_{xy}\left(B_x^2+B_y^2\right)+\chi_z\,B_z^2\right],
    \end{equation}    
    where $\chi_{ab}$ and $\chi_z$ are the diamagnetic susceptibility of the graphite sheet in the $x-y$ plane and $z$ direction respectively, and $B_q$ is the magnetic field strength along the axis $q\in x,y,z$. Here we have assumed that the thickness of the graphite sheet $\delta$ is sufficiently thin, so that the magnetic field at the top and bottom of the sheet is of equal value. This potential only becomes trapping once we introduce the linearized gravitational potential $mg\,z$, as such the total trapping potential is given: 
    \begin{equation}
        U(x,y,z)= m g\,z + U_{\mathrm{mag}}(x,y,z).
        \label{eq:Utotal}
    \end{equation}    
    The spring constants are easily extracted via the second derivatives,    
    \begin{equation}
        k_q=\left.\frac{\partial^2 U}{\partial q^2}\right|_{\rm q_{eq}},
        \label{eq:spring_constants}
    \end{equation}
    where $q\in(x,y,z)$, and $q_{eq}$ denotes the equilibrium positions of $\partial U/\partial q=0$.
    From the simulations, we find that $k_z \approx 51k_{x,y}$, so while our optical detection is less sensitive to these components, our sensor is actually displaced more under forces along these axes. 
    We note that by varying the diamagnetic susceptibility $\chi\to\chi'$ and the  magnetic remanence $B_r\to B_r'$, the magnetic potential changes accordingly
    \begin{equation}
       U_{mag}\to \left(\dfrac{\chi'}{\chi}\right)\left(\dfrac{B_r'}{B_r}\right)^2U_{mag}. \label{U_scale}
    \end{equation}
 We see that changing $B_r$ or $\chi$ changes the strength of the trapping potential, which would also lead to a shift in the equilibrium positions. Varying $g$ (or equivalently $m$) on the other hand, keeps the shape of magnetic potential fixed, but directly changes the equilibrium position of the oscillator.\\ 

\paragraph{\textit{Optimizing Sensitivity ---}}
 From equations~\ref{eq:thermalnoise},~\ref{eq:detectionnoise} and Hooke's law, it is clear that the smaller spring constant $k$ leads to a larger acceleration sensitivity. To optimize the sensitivity of our sensor, we thus want to minimize this quantity. Since the height of levitation also changes when $B_r$ and $\chi$ change, the actual scaling of $k_z$ is somewhat different from that shown in equation~\ref{U_scale}. From our simulations, we find that the spring constant along our principal axis, $k_z$, scales with the strength of the magnetic array as $k_z \propto -\,B_{r}^3$ and equally, $ k_z \propto -\chi^3$, as shown in fig.~\ref{fig:springstiffness}A. and ~\ref{fig:springstiffness}B., which we discuss further in Supp.~C. 
To minimize the spring constant in $z$, stronger magnets and greater magnetic susceptibilities can be used. We used a \SI{1.48}{T} N52 magnet of size \SI{25.4}{mm} $\times$ \SI{25.4}{mm} $\times$ \SI{12.7}{mm} which provided the highest remanence readily available. HOPG was used due to its high magnetic susceptibility and low density, making it an ideal candidate for the levitated test mass at room temperature. 
While the optimization of the $z$ spring constant is our main approach to improving the sensor, other improvements, such as further vibration isolation are commonly implemented in other gravimeters. Any improvement to the interferometric readout of the sensor is captured in equation~\ref{eq:detectionnoise}. Currently, the detection noise is completely eclipsed by our signal and other noise sources at \SI{0.3}{\micro Gal/\sqrt{Hz}}. 
As the $z$ spring constant decreases, the equilibrium height is increased and the axial confinement in $x$ and $y$ decreases rapidly, as can be seen if Fig.~\ref{fig:springstiffness}D. For incremental improvements, such as stronger permanent magnets, the axial trapping remains stable. However, for drastic improvements the geometry would need to be altered to maintain axial stability.\\

\paragraph{\textit{System Stability ---}}
The scaling of the system can also be used to analyze the stability of the system, as changes to the spring constant caused by the model parameters $B_r$, $\chi$, $g$ and $T$ directly influence the measured acceleration. Fig.~\ref{fig:springstiffness} shows the stability of the system as a function of its parameters. Varying the magnetic remanence $B_r$ by approximately 10\% leads to a 8.3\% change in $k_z$. Varying the diamagnetic susceptibility $\chi$ by 10\% leads to a change of approximately 3\% of the spring constant, a factor 3 smaller than the magnetic field scaling. These scalings are shown in Fig.~\ref{fig:springstiffness}A. and B.\\ 
The remnant magnetization and the diamagnetic susceptibility have further temperature dependence, which is shown in Fig.~\ref{fig:springstiffness}C. 
The remnant magnetization $B_r$ is widely reported by industrial suppliers to change by at most \SI{-0.12}{\%/\celsius} while below \SI{100}{\celsius}. 
The diamagnetic susceptibility of HOPG decreases for increase in temperature, by \SI{-0.15}{\%/\celsius} \cite{magnetochemistry4040052}. These scalings all contribute only to a minimal change to the stiffness of the system as the temperature changes, namely \textless~\SI{1}{\%} for a temperature fluctuation of approximately $5\celsius$, showing a high degree of stability when compared to silicon based MEMS gravimeters \cite{Middlemiss2016}.

Over variations in local gravity in the order of 20 times the semi-diurnal lunar tides, the $z$ spring constant varies by less that 1 ppm. This corresponds to the frequency changing by less than 0.5 ppm, or the acceleration sensitivity changing by less than 1 ppm, and does not substantially change the dynamics of the system. For larger variations in local gravity, the $z$ spring constant varies quadratically with the change in local gravitational acceleration. As such, the sensitivity will decrease quadratically under high accelerations, but increase quadratically under low accelerations. This introduces the possibility of conducting highly sensitive experiments in a microgravity environment such as a drop-tower or a satellite in a earth-bound free-falling orbit. LOMS has already been demonstrated in such environments \cite{Homans_2025}.
If the local gravitational acceleration increases beyond $3.14~g$, the sensor discussed in this work is forced into the magnetic array, providing an upper limit to the acceleration measured. Another fundamental limit is found in the thermal demagnetization of the levitating array. For the sensor discussed in this paper, the remanent field of the magnetic array must be larger than \SI{0.844}{T}, below which Earth's gravitational acceleration will force the graphite into contact with the magnetic array. In practice, the magnetic field of the array is only reduced below this value beyond the Curie temperature (\SI{310}{\celsius} for the magnets used in this sensor), at which point the magnets are fully demagnetized. These limits are dependent on the shape of the levitated sheet, and as such can be optimized for.
    
\begin{figure*}[t!]
        \centering    
        \includegraphics[width=2\columnwidth]{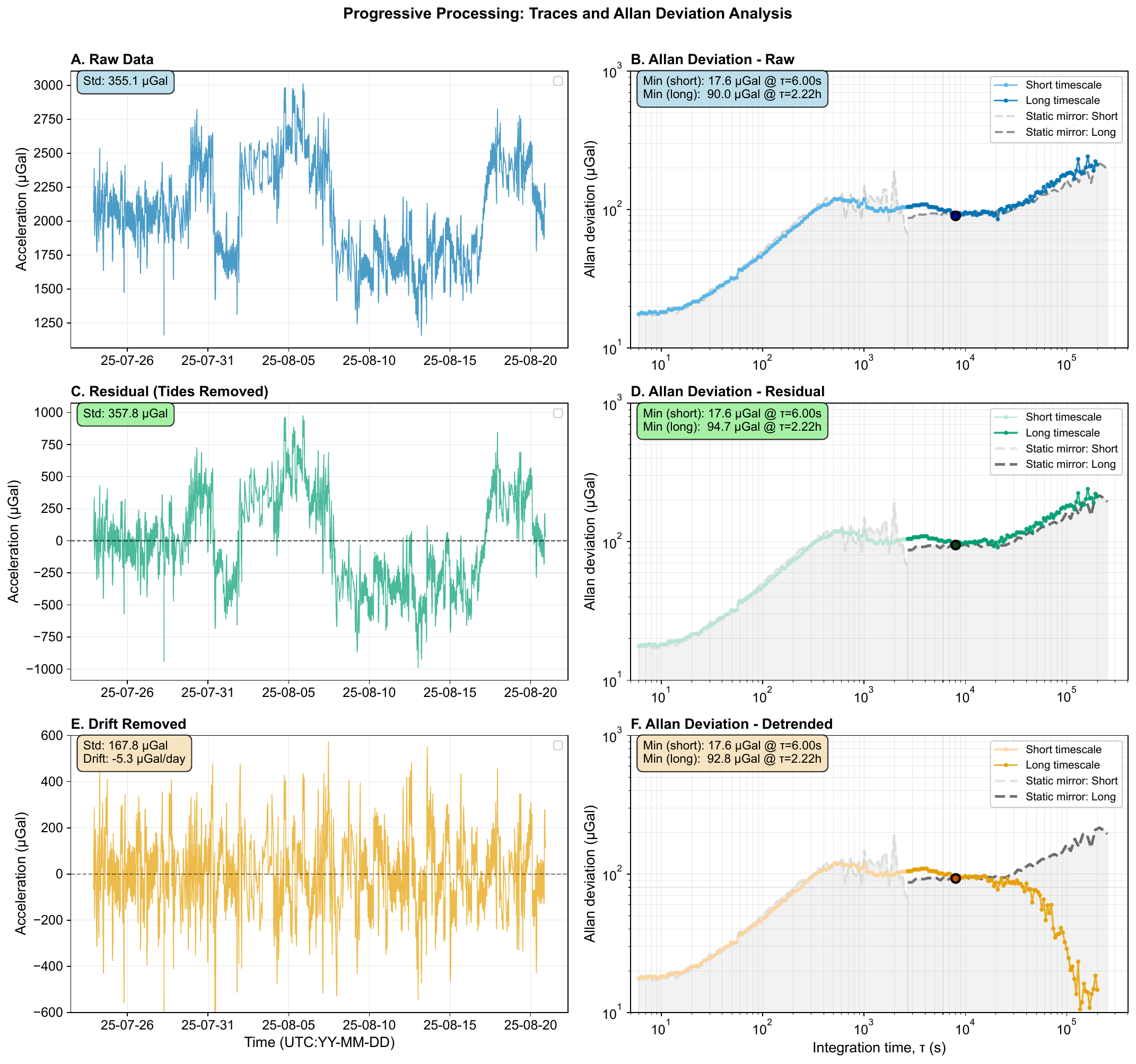}
        \caption{\textbf{Temporal Stability of the Sensor, shown in time-traces and Allan deviation ---} Two minute binned traces and Allan deviations: 28 day data set (as used in Fig.~\ref{fig:TimeTrace}) time traces plotted using: \textbf{A.} raw data, \textbf{C.} Model identified tides residual, \textbf{E.} Tides residual with long term drift correction with corresponding Allan deviations \textbf{B.},\textbf{D.} and \textbf{F.} respectively. The dashed black line represents the instrumental noise floor of the setup with a static mirror in place of the levitated test mass. The electrical noise floor of the detection system is well below the range of accelerations represented here so it has been omitted due to it not being a limiting factor. For reference the electrical noise is consistently below \SI{0.7}{\micro Gal} over the range shown.}
        \label{fig:AllanDeviation}
  \end{figure*}

\paragraph{\textit{Allan Deviation ---}}
We extract Allan deviations from our time-traces to determine the time-dependent sensitivity of LOMS. Fig.~\ref{fig:AllanDeviation} shows the Allan deviation of our measured acceleration for integration times from \SI{1}{s} to \SI{3}{days} at different steps in our data analysis. For short timescales, from \SI{6}{s} up to around \SI{20}{s}, the system experiences minimal drift and stays around a resolution of \SI{18}{\micro Gal}. From equations~\ref{eq:thermalnoise} and \ref{eq:detectionnoise}, we extract that the thermal noise limit at \SI{300}{K} of the sensor is \SI{58.5}{n Gal/\sqrt{ Hz}}. We further extract the detection noise to be \SI{194}{n Gal/\sqrt{Hz}}~\footnote{For a 16 bit effective number of bits (ENOB) limited detection.}. When these figures are integrated over the same \SI{6}{s} acquisition time they can be compared directly to the Allan deviation, with a thermal limit at \SI{24}{nGal} and detection limit at \SI{97}{nGal}, respectively. This resolution is competitive with the state-of-the-art in MEMS and atom interferometer gravimeters and is achievable over timescales of several seconds, widening the range of possible applications. Over the period of one day, the system drifts by \textless \SI{100}{\micro Gal}, which is also competitive with MEMS gravimeters.
The limit of the stability of the current system is due to environmental factors such as seismic noise, thermal drifts and the stability of the off-the-shelf optics used. From the good agreement between raw-data Allan deviation and the fixed-mirror Allan deviation, we conclude that the majority of the drift and instability is caused by shifts in the readout system, rather than any drift in the levitated mass.
Our gravimeter was not designed to mitigate the seismic and thermal environment in order to showcase the stability of the sensor. While the sensor was placed on an active anti-vibration platform, this platform had a lowest working frequency of \SI{0.5}{Hz}, providing minimal to no vibration isolation in the frequency range shown in Fig.~\ref{fig:AllanDeviation}. 

\paragraph{\textbf{Conclusions ---}}
\label{sec:conc}

We detect the lunar and solar tides using diamagnetic levitation and using strictly low cost commercial-of-the-shelf components, showcasing their potential sensitivity to gravitational and seismic signals. Notably, the sensor offers full 3D gravitational acceleration sensitivity when compared to conventional state of the art mass-on-spring gravimeters, which can be extracted by spatially resolved quadrant photodiode detection.  The small form factor (\SI{50 }{ml}), low power consumption (\SI{2}{W}), and low cost and complexity make the sensor well-suited for mass adoption, remote application, and space-based applications. Typically, the size and power consumption make common state-of-the-art gravimeters ill suited for space based measurements. Current satellite based gravimetry, such as GRACE, GOCE, and LISA-PF, all utilize free falling masses and cost hundreds of millions. In contrast, our LOMS device demonstrated here has the potential to take gravitational data in orbit for a fraction of the cost, allowing for scaling-up to arrays of sensing satellites. Compared with the current state-of-the-art in gravimetry, our sensor functions at room temperature without the need for expensive and noisy cooling solutions. The long term stability of the sensor can be readily improved by engineering a more stable mounting of the optical detection system.

The currently achieved sensitivity offers the detection of a variety objects of interest. By approximating features as horizontal cylinders beneath the surface, a standard two lane car tunnel can be detected up to a maximum depth of \SI{350}{m} and an analogous mineral deposit to the one in the Kiruna mine could be detected down to a depth of \SI{6.8}{km}, both far exceeding the features real world depth. For features closer than their maximum detectable depth, their shapes can be mapped in addition.

\paragraph{\textbf{Outlook ---}}
\label{sec:out}
To access the full 3D sensitivity of our LOMS device, additional modeling must be done to account for the difference in spring constant of the different modes and the angular deflection caused by the shape of the potential when the sensor is accelerated in the $x-y$ plane. To reject common-mode noise and drift, the next iteration of the sensor is planned to include three additional levitated masses in a gradiometric configuration. By spacing identical sensors along the $x-, y-$ and $z-$axis, the combined sensor is able to spatially resolve the mass and distance of over-densities while rejecting readout noise and vibrational noise. A straightforward improvement to our device sensitivity can be made by switching to grade N56 magnets, offering a $7\%$ increase in remanence, increasing sensitivity by $22\%$. While superconducting test masses used in state-of-the-art gravimeters are perfect diamagnets, they require cryogenic temperatures. The solid HOPG used in our device has excellent diamagnetic behavior, but is also a good conductor. This quality causes higher eddy-current damping when our sensor is displaced, reducing the Q factor. Whilst thermal noise of the sensor was shown not to be a limiting factor to our device in the current state, research has been done on reducing this damping by altering the graphite test mass \cite{GraphiteCut,GraphiteComposite1,graphitecomposite2}. 
From our theoretical model, we find that a small reduction in the spring constant along z can cause a massive reduction to our spring constant along x and y, positioning our LOMS device to be extremely sensitive to in-plane accelerations, at the cost of a lower sampling rate. Since most seismic effects of interest are far below our current sampling rate, this is an effective way to increase the sensitivity of our sensor. 

\mysection{Acknowledgments}
TMF, ES, MH and HU acknowledge funding from the EU Horizon Europe EIC Pathfinder project QuCoM (10032223), from the UK funding agency EPSRC (grants  EP/V035975/1, EP/V000624/1, EP/W007444/1, EP/X009491/1), and from the Leverhulme Trust (RPG-2022-57), as well as support from the QuantERA II Programme (project LEMAQUME) that has received funding from the European Union’s Horizon 2020 research and innovation programme under Grant Agreement No 101017733. FJH acknowledges the EU EIC Pathfinder project QuCoM (101046973). 

\bibliography{biblio}


\clearpage
\newpage

\onecolumngrid

\begin{center}
\large
\textbf{
    \textit{Supplementary Information for} \\ Detection of Lunar Tides by Diamagnetic Levitation
    }

\end{center}

\setcounter{figure}{0} 
\setcounter{equation}{0} 
\renewcommand{\figurename}{\bf Fig.} 
\renewcommand{\thefigure}{S\arabic{figure}} 

\renewcommand{\theequation}{S\arabic{equation}}

\noindent
The following is supplementary information for the paper {\it Detection of Lunar Tides by Diamagnetic Levitation}. We give more details on the A. Conversion from detected voltage to tidal acceleration, the B. Spectral filtering of the signal, and the C. Analytical model to describe the motion of the graphote probe mass in the diamagnetic trap.  We also give a table of all relevant baseline parameters used to model the gravimeter device.

\section{A.~Conversion from detected voltage to tidal acceleration}
\label{sec:Conversion}

\noindent
Far below resonance, the force sensitivity of a mechanical resonator along an axis $q$ is given by the spring constant of the mode under test, $k_q$, which gives the conversion between the measured displacement and the force using Hooke's law, $F_q = k_q\,q$. The natural frequencies of the resonators modes can be used to determine the spring constant from $k_q = m\omega_{n,q}^2$ where $m$ is the mass of the levitated sheet and $\omega_{n,q}$ is the natural angular frequency.     
As the local gravitational field changes due to tidal forces, the test mass undergoes vertical displacement, which is detected through optical interferometry. The interferometer outputs two quadrature channels (sine and cosine components) that encode the displacement information as intensity changes in the interference pattern. The gravimeter functions as a simple harmonic oscillator with natural frequency $f_n$. 
For such a system, the relationship between the displacement $x(t)$ and the acceleration $a(t)$ is given by
    \begin{equation}
        a(t) = -\omega_n^2 x(t) = -\left( 2\pi f_n \right)^2 x(t),   
    \end{equation}
where $\omega_n$ is the angular frequency. 
The resonant frequency $f_n$ was determined from the position data power spectral density, 
and the measured displacements were converted to acceleration using this relationship.
    
\begin{figure}
     \centering
     \begin{subfigure}[a]{\columnwidth}
        \centering
        \includegraphics[width=0.75\columnwidth]{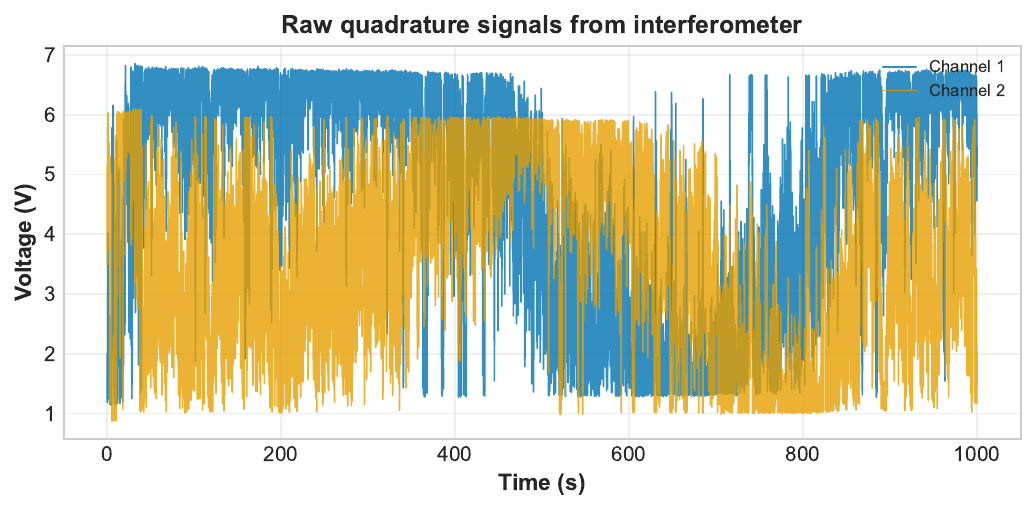}
        \caption{Plot of raw trace file pair from the two channels recorded via oscilloscope}
        \label{fig:S1a}
     \end{subfigure}
     \begin{subfigure}[b]{\columnwidth}
        \centering
        \includegraphics[width=0.5\columnwidth]{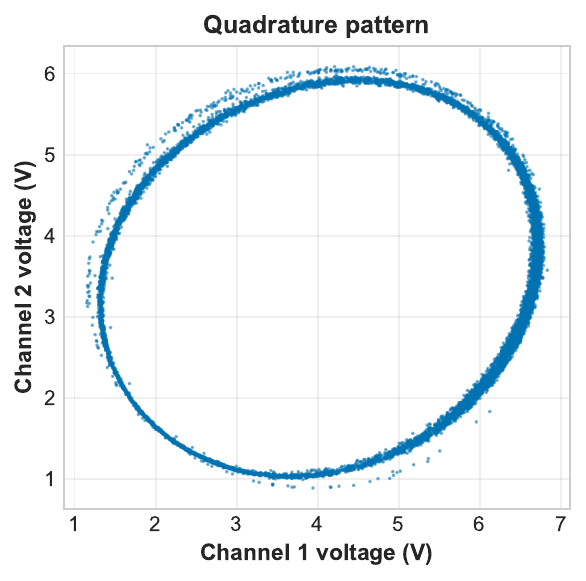}
        \caption{Corresponding quadrature plot produced from file pair prior to correction}
        \label{fig:S1b}
    \end{subfigure}
        \caption{Example of the first processing step using a single chunk of a \SI{1000}{s} of data, combining two sample trace files to produce the desired quadrature.}
        \label{fig:S1}
    \end{figure}

Raw oscilloscope trace files were loaded in pairs according to their naming convention and ordered chronologically by file creation time. Each pair consists of two quadrature channels that together encode the interferometric phase information, an example of which is shown in Figure~\ref{fig:S1a}. Channel symmetry was verified to ensure complete data sets, and an initial down sampling of a factor 10 was applied to manage memory requirements during processing.
    
When the two channels are plotted against each other in IQ space, they produce a quadrature signal ideally represented as a circle, as shown in Figure~\ref{fig:S1b}. However, raw quadrature plots deviate from a perfect circle due to differing gains in the separate detectors, imperfection of the incident polarization and slight imbalances of the Polarizing Beam Splitters (PBS) splitting, leading to ellipticity. Further, the circle formed is not centered around the origin since interferometric intensity is strictly positive. A further offset of the quadrature is due to amplitude imbalances between the interferometer arms (non-unity fringe visibility) and DC offsets in the detectors. To center this ellipse at the origin, the ellipse parameters are determined by solving a generalized eigenvalue problem. The geometric center is then extracted from these parameters, and the data is shifted to place this center at (0,0). This ellipse-based approach avoids the bias that would result from simple mean-averaging, which assumes circular symmetry and equal weighting around the circumference.
   
\begin{figure}
     \centering
     \begin{subfigure}[a]{\columnwidth}
        \centering
        \includegraphics[width=0.5\columnwidth]{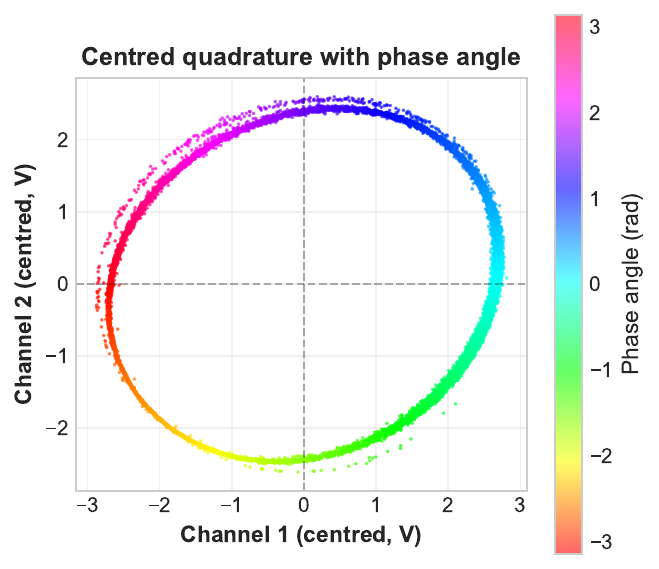}
        \caption{Plot of centered quadrature, enabling proper angle extraction.}
        \label{fig:S2a}
     \end{subfigure}
     \begin{subfigure}[b]{\columnwidth}
        \centering
        \includegraphics[width=0.75\columnwidth]{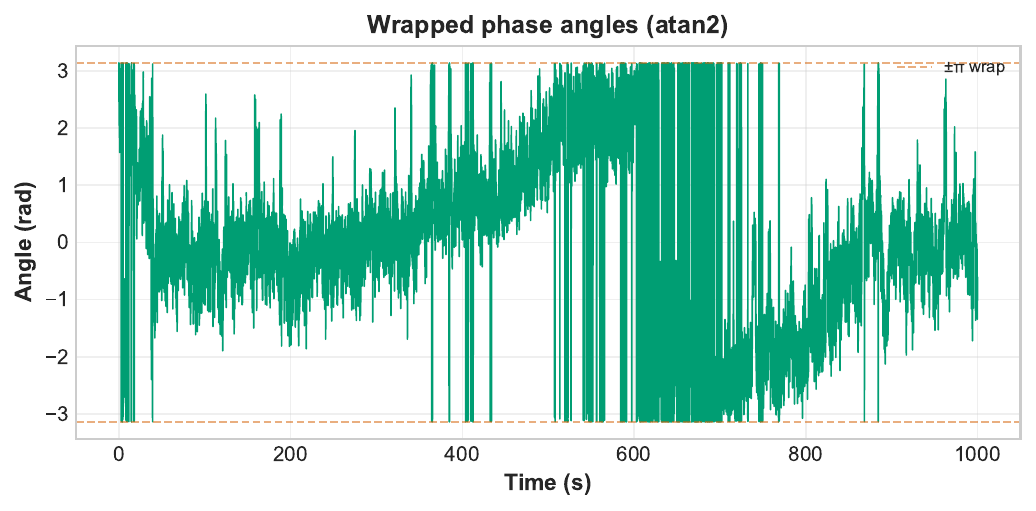}
        \caption{Wrapped phase trace extracted from corrected quadrature.}
        \label{fig:S2b}
     \end{subfigure}
     \begin{subfigure}[c]{0.75\columnwidth}
        \centering
        \includegraphics[width=\columnwidth]{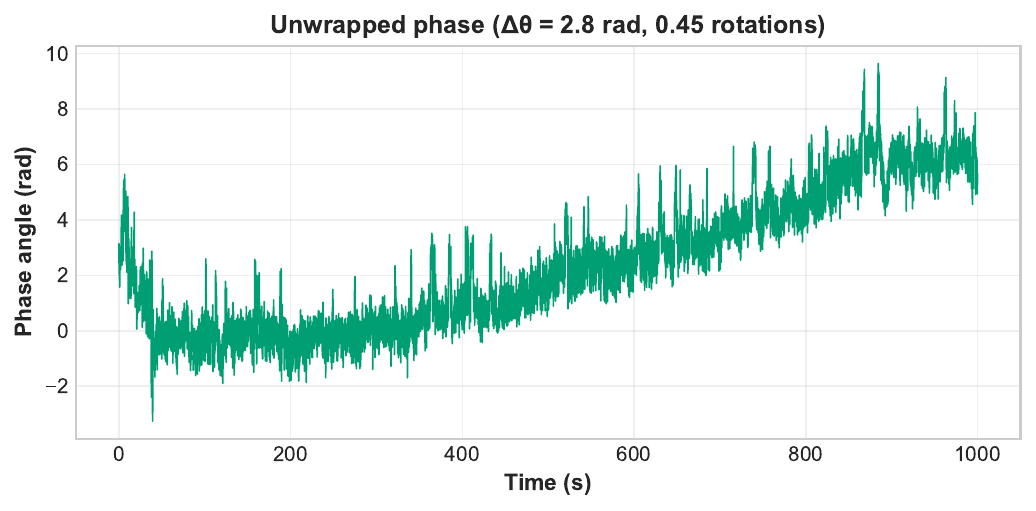}
        \caption{Unwrapped phase data.}
        \label{fig:S2c}
     \end{subfigure}
     \caption{Example of second main processing step, angle changes can be determined from the corrected quadrature and unwrapped into a continuous trace. This was produced using a sample file pair from the main data set.}
     \label{fig:S2}
\end{figure}

Angular positions were extracted from the corrected quadrature plots (Figure~\ref{fig:S2a}) using the \texttt{arctan2} function, which uniquely determines the angle for each point in the quadrature plane. The resulting angle segments were differentiated (using \texttt{np.diff}) to produce relative angle changes between consecutive measurements, producing the wrapped phase trace shown in Figure~\ref{fig:S2b}. At segment boundaries, the algorithm assumes the shortest angular path to handle phase wrapping correctly. The resulting wrapped phase is then unwrapped into a continuous trace as shown in Figure~\ref{fig:S2c}.

\begin{figure}
    \centering
     \begin{subfigure}[a]{\columnwidth}
        \centering
        \includegraphics[width=0.5\columnwidth]{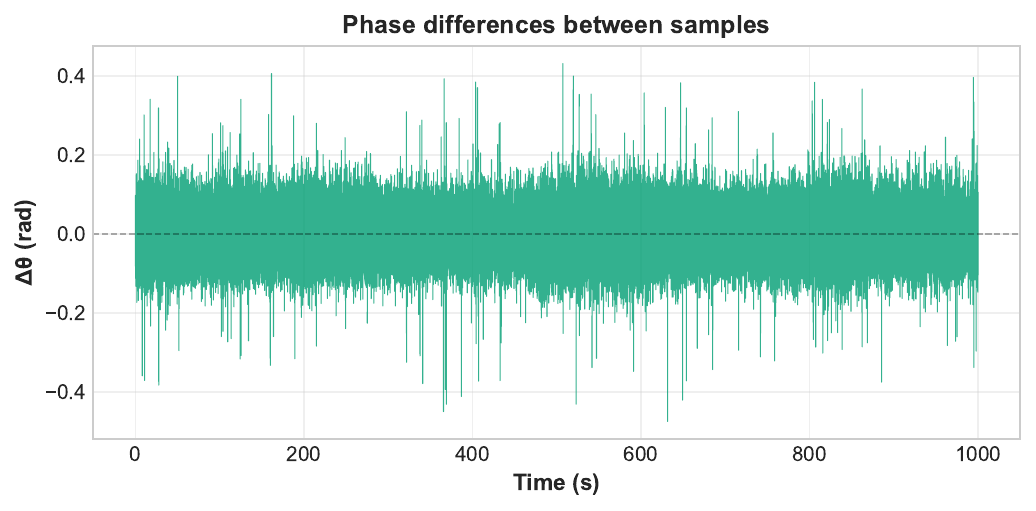}
        \caption{Plot of angle changes $\Delta\theta$(t) from the unwrapped phase.}
        \label{fig:S3a}
     \end{subfigure}
     \begin{subfigure}[b]{\columnwidth}
        \centering
        \includegraphics[width=0.5\columnwidth]{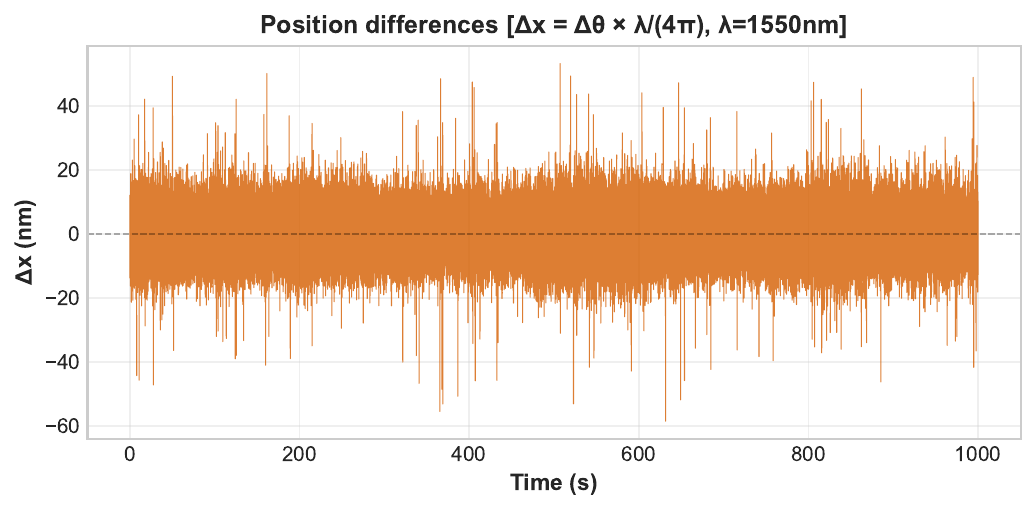}
        \caption{Plot of position changes $\Delta$x(t) produced via application of conversion factor.}
        \label{fig:S3b}
     \end{subfigure}
     \begin{subfigure}[c]{\columnwidth}
        \centering
        \includegraphics[width=0.5\columnwidth]{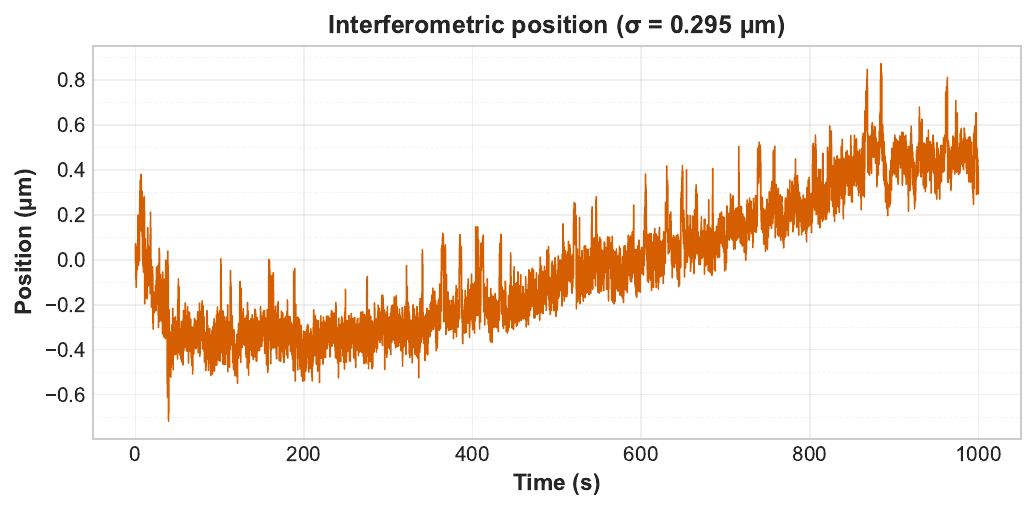}
        \caption{Integrated position changes provide a trace of the oscillator position x(t).}
        \label{fig:S3c}
     \end{subfigure}
        \label{fig:S3a_fig}
     \begin{subfigure}[d]{\columnwidth}
        \centering
        \includegraphics[width=0.5\columnwidth]{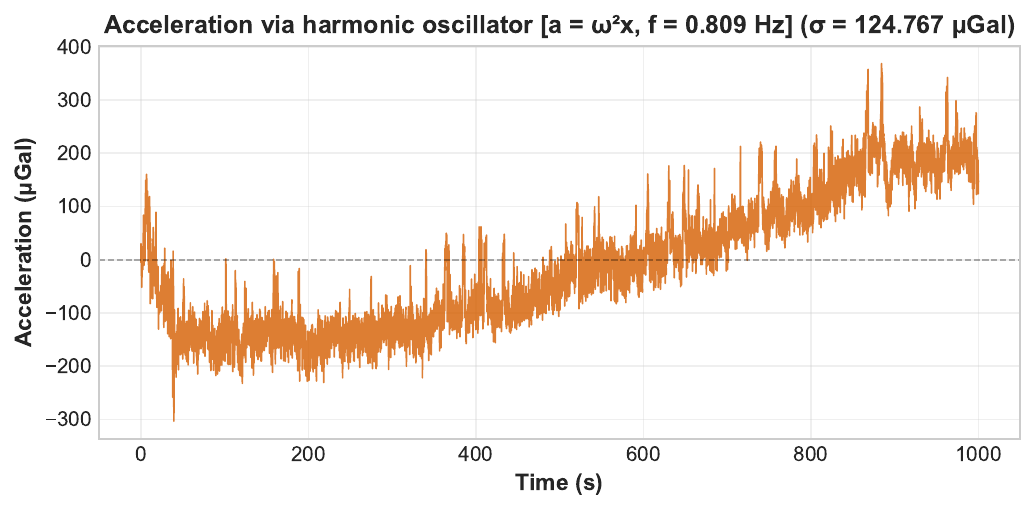}
        \caption{Acceleration trace obtained via application of the flat transfer specified in equation one of the main text to the position trace.}
        \label{fig:S3d}
     \end{subfigure}
    \caption{Example of third main processing step (cont.): application of conversion factor prior to integration. It is possible to obtain a trace of the position change $\Delta$x(t), which can trivially be integrated to obtain the desired position trace as in (c.).}
    \label{fig:S3b_fig}
\end{figure}

\noindent
The relationship between phase change and physical displacement in an interferometer is given by
    \begin{equation}
        \Delta X = \frac{\lambda}{2} \cdot \frac{\Delta\theta}{2\pi},
    \end{equation}
where $\lambda$ is the optical wavelength of the source laser, $\Delta\theta$ is the phase change in radians, and $\Delta X$ is the corresponding physical displacement. The factor of $\lambda/2$ arises because a full round-trip path through the interferometer (displacement and return) accumulates twice the phase change of a single pass. Simplifying this relationship yields equation \ref{dxdtheta}.
    \begin{equation}
        \Delta X = \frac{\lambda}{4\pi} \cdot \Delta\theta
    \label{dxdtheta}
    \end{equation}
This conversion factor, $\lambda/(4\pi)$, was applied to all angle changes to construct a continuous trace of position changes in units of meters, shown in Figure~\ref{fig:S3a} and Figure~\ref{fig:S3b} respectively. These position changes are then integrated to obtain the oscillator position trace $x(t)$ shown in Figure~\ref{fig:S3c}, from which the acceleration trace (Figure~\ref{fig:S3d}) is obtained via application of the flat transfer function specified in equation 1 of the main text.\\
 \raggedbottom

\noindent
    Lunar tidal signals have characteristic periods on the order of 12.42 hours (semidiurnal tide M2) and longer, corresponding to frequencies below approximately \SI{2.2e-5}{Hz}. The raw sampling rate of the gravimeter far exceeds the Nyquist requirement for these signals, making down-sampling both practical and necessary for computational efficiency. A half-hour binning procedure was implemented, in which position data falling within each 30-minute window was averaged to produce a single value at the bin center. This procedure serves a dual purpose: It acts as a low-pass filter that attenuates high-frequency noise and environmental vibrations while preserving the tidal signals of interest, and dramatically reduces data volume.

    To assess the gravimeter's response to known gravitational signals, the measured data was compared against theoretical tidal predictions. The position data was converted to acceleration data via the flat transfer function specified in equation 1 of the main text, 
    \begin{equation*}
        a_q = (2\pi f_{n,q})^2~q    
    \end{equation*}
    and converted to units of \si{\mu Gal}. This was exported in time series format (.tsf) along with corresponding timestamps and coordinates for the experiment. Tsoft software was used to generate a comprehensive theoretical prediction from the WDD Solid Earth model \cite{dehant1999tides} with ocean loading correction parameters produced from the FES2014b model\cite{Carrere2016FES2014,lyard2021}, with the included tidal components at the instrument location and observation times noted from the subsequent model parameters. 

    The traces observed in Figure 2 of the main text are simply the contents of the observed data within the filter bands specified in section B plotted in the time domain against the raw data trace and acceleration trace predicted by the model. The Sun-Moon alignment trace was created using a reference sturgeon moon (full moon) date obtained from the Royal Museums Greenwich website \cite{fullmoon2025} and the period of the lunar month/phases. This gives an independent reference for the solar–lunar alignment. Although the data contains additional beating patterns with periods close to the spring–neap cycle, comparison of these patterns with a reconstruction of the known solar–lunar alignment showed a disagreement in phase. This independent verification prevents us from incorrectly identifying coincidental beating patterns as the spring–neap tide, which would be inappropriate given the complexity of modelling three‑dimensional gravitational vectors.
    
\newpage

\section{B.~Spectral filtering of the signal}
\label{sec:Filter}
\noindent
The vibration isolation stage used to isolate our sensor from seismic noise has a lowest effective filtering frequency of 0.5 Hz. Since the tidal signal is far below this frequency, the filtering performed by the stage is minimal in this spectral band. To this end, we made use of spectral filtering to isolate the tidal components affecting our sensor. In this section we discuss the filtering used and verify that the filter does not create a spurious signal.

\mysection{i.~Construction of the Filter}
\noindent
Our filter design is based on constructing a characteristic spectral fingerprint derived from the Tsoft tidal model for our geographic location. The design process is as follows: we compute the power spectral density (PSD) of the Tsoft model and identify prominent spectral peaks within the tidal band (6–50 hour periods). Peaks are identified by their amplitude relative to the local spectral background, with a threshold set at 30 percent of the maximum spectral amplitude in the tidal band.
Each identified peak is subsequently assigned to a known tidal constituent by comparing its period to the expected periods from tidal theory. A tolerance of ±0.4 hours is used to account for finite frequency resolution.

\noindent
The major constituents considered are the principal lunar semidiurnal M2 (12.42 h), the principal solar semidiurnal S2 (12.00 h), the larger lunar elliptic semidiurnal N2 (12.66 h), the combined lunar-solar declinational and solar diurnal K1+P1 respectively (24.00 h), the principal lunar diurnal O1 (25.82 h), and the larger lunar elliptic diurnal Q1 (26.87 h).
For each identified constituent, a narrow pass-band is created centered on the constituent's frequency with a bandwidth of ±0.2 hours in period space (±0.15 hours for semidiurnal constituents, ±0.25 hours for the combined K1+P1 peak). This bandwidth is sufficient to capture the constituent's spectral power while minimizing contamination from adjacent frequencies.
A notable consideration in the design surrounds the K1 and P1 constituents: Due to their close proximity in frequency space (23.93 h and 24.07 h respectively, separated by only 0.14 hours), K1 and P1 cannot be resolved separately given our measurement duration. We therefore treat them as a combined constituent (K1+P1) centered at their average period of 24.00 h, with an expanded pass-band. 

\noindent
Additionally, the shallow-water constituent MK3 (8.18 h) represents a minor contribution in the one-dimensional tidal model spectrum, we observed an amplified MK3 component in our data, likely due to local hydrological effects. A manual band was specified to ensure this constituent could be passed.

\noindent
The filter is implemented as a frequency-domain brick-wall (rectangular) mask  meaning a filter response of one within pass-bands and a response of zero with no tapering.
This implementation is critical for our validation approach: the filter applies no amplitude modification within pass-bands; constituent amplitudes are preserved exactly as they appear in the data. Only frequencies outside the defined pass-bands are removed. This ensures that any amplitude ratios measured between constituents reflect genuine relationships present in the data, rather than artifacts introduced by filter shaping or scaling.

\noindent
The filtering procedure is applied in the frequency domain via the Fast Fourier Transform (FFT). The procedure consists of computation of the FFT of the mean-centered time series, application of the filter mask H(f) to both positive and negative frequencies and lastly computation of the inverse FFT to obtain the filtered time series.

\begin{figure}
    \centering
    \begin{subfigure}[a]{\columnwidth}
        \centering
        \includegraphics[width=0.35\columnwidth]{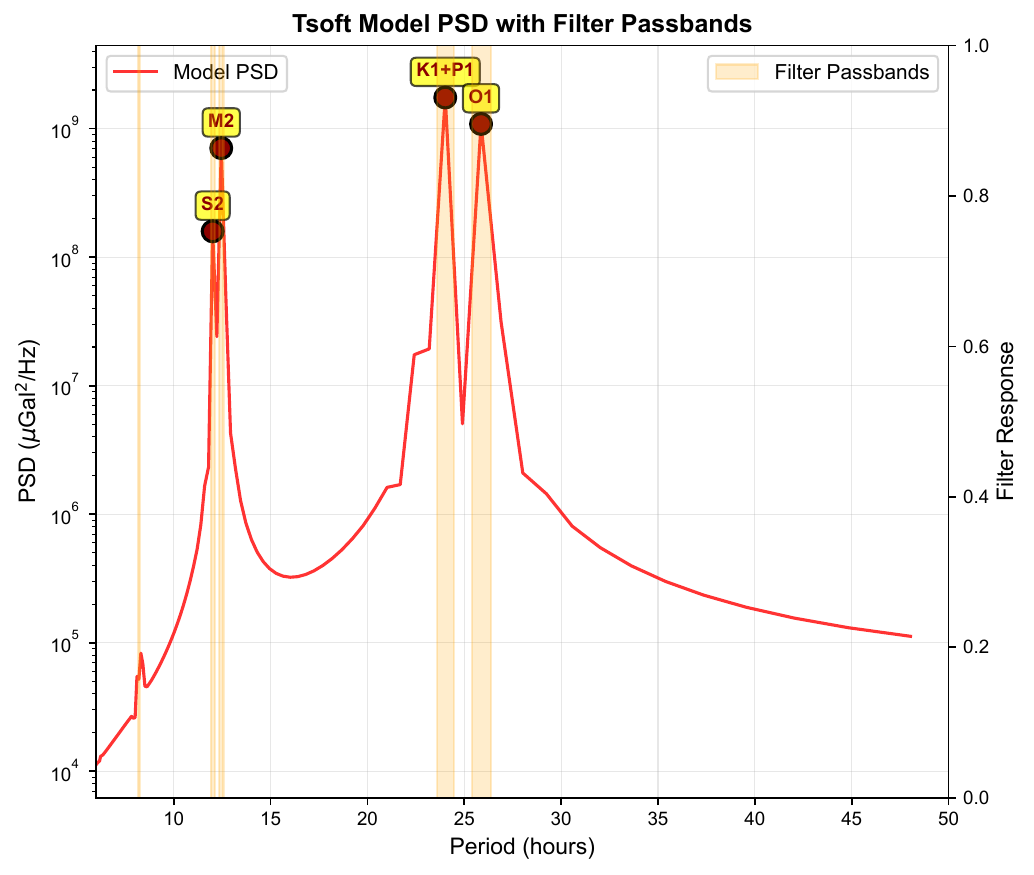}
        \caption{Tsoft model PSD with identified peaks and filter pass-bands overlaid.}
        \label{fig:S4a}
    \end{subfigure}
    \label{fig:S4}
    \begin{subfigure}[b]{\columnwidth}
        \centering
        \includegraphics[width=0.35\columnwidth]{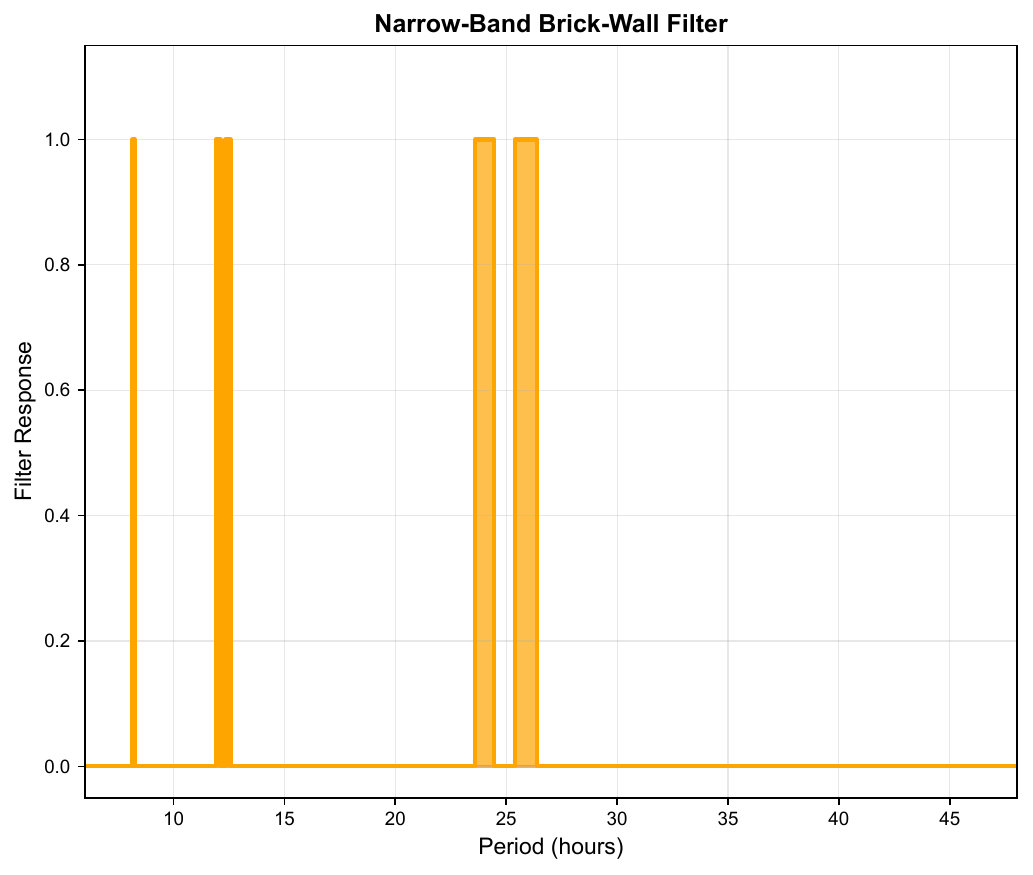}
        \caption{Filter response function showing the narrow-band brick-wall structure.}
        \label{fig:S4b}
    \end{subfigure}
    \begin{subfigure}[c]{\columnwidth}
        \centering
        \includegraphics[width=0.35\columnwidth]{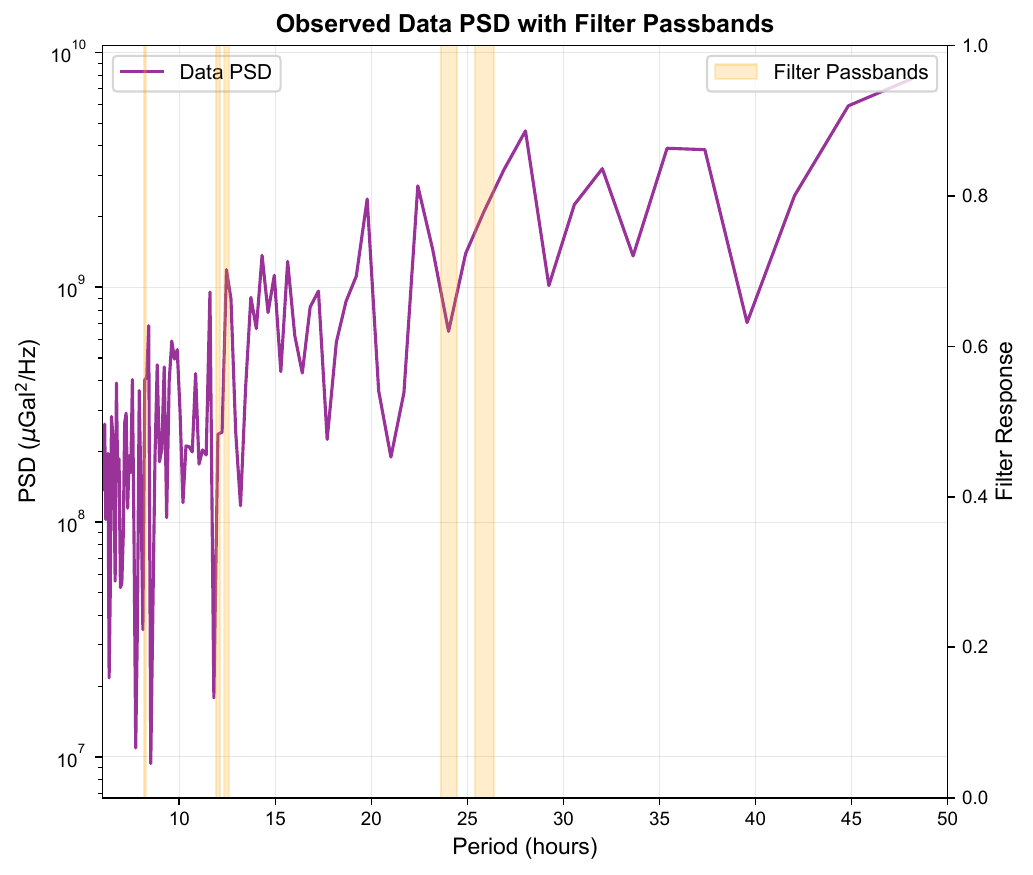}
        \caption{Observed data PSD with filter pass-bands overlaid.}
        \label{fig:S4c}
    \end{subfigure}
    \begin{subfigure}[d]{\columnwidth}
        \centering
        \includegraphics[width=0.35\columnwidth]{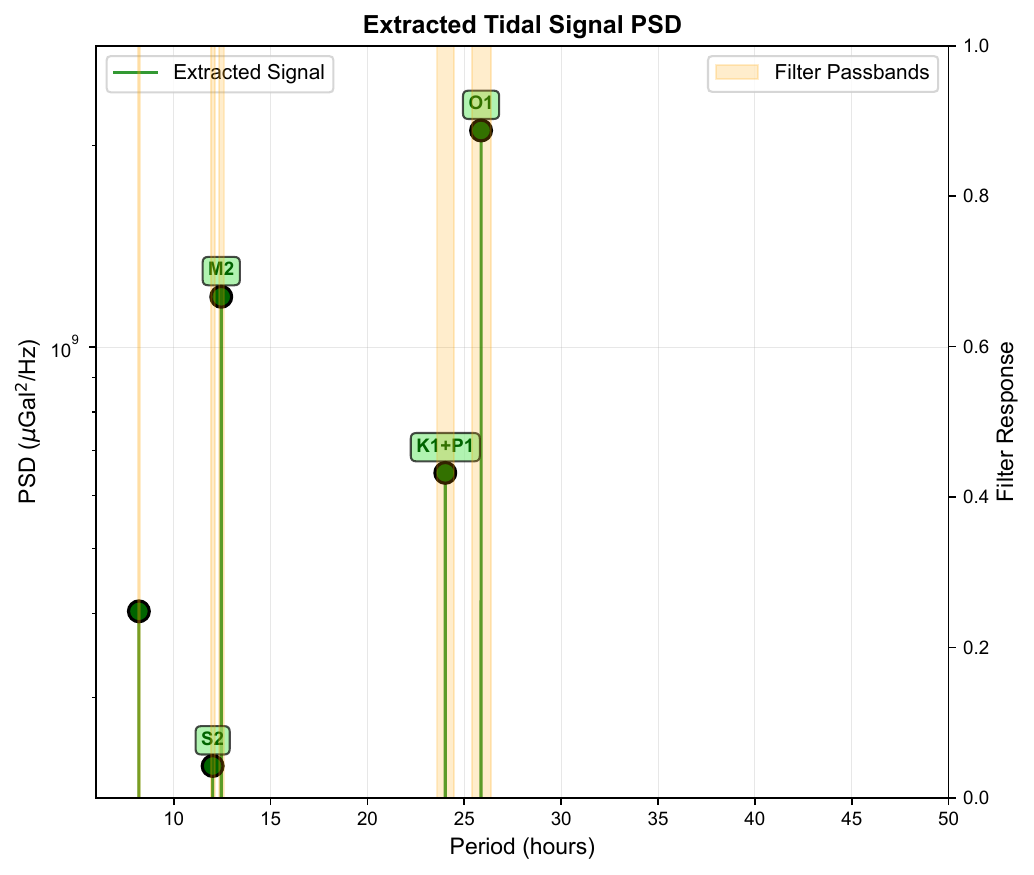}
        \caption{Extracted tidal signal PSD after filtering.}
        \label{fig:S4d}
    \end{subfigure}
    \caption{Filter construction process showing (a) the Tsoft model PSD with identified peaks and filter pass-bands overlaid, (b) the filter response function showing the narrow-band brick-wall structure, (c) the observed data PSD with filter pass-bands, and (d) the extracted tidal signal PSD after filtering.}
    \label{fig:S4_cont}
\end{figure}

\mysection{ii.~Verification of the Filter Response Through Amplitude Ratios}
A critical concern with any spectral filtering approach is whether the filter itself could create spurious signals or artificially generate correlations that appear to validate the presence of tidal constituents. We address this through a comprehensive validation scheme based on amplitude ratio analysis.

The relative amplitudes of tidal constituents are determined from the Wahr-Dehant-Defraigne (WDD) solid earth tidal model with ocean loading parameters from the FES2014b tidal model and depend on the sensor's geographic location. For our location, the Tsoft model predicts specific ratios between major constituents (e.g., M2/S2, M2/O1, K1+P1/O1). If our filter successfully extracts genuine tidal signals, the amplitude ratios measured in the filtered data should closely match those predicted by the model.

\begin{figure}[htbp]
    \centering
    \includegraphics[width=0.75\textwidth]{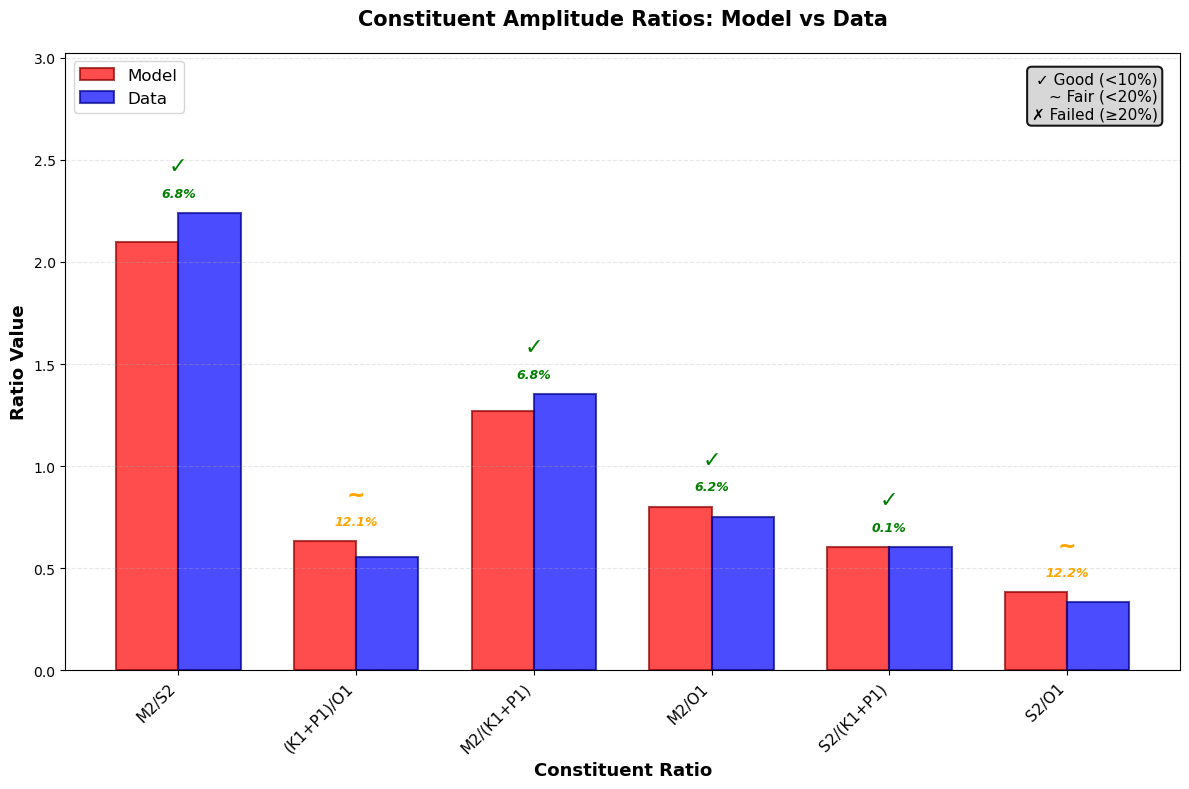}
    \caption{Constituent Amplitude Ratios: Model vs Data: With the Tsoft model ratios (red) and the observed acceleration data ratios(blue). Arbitrary tiered thresholds were created to grade matches. The agreement between our filtered data and the Tsoft model is Fair or better across all ratios tested.}
    \label{fig:overall}
\end{figure}

We extract constituent amplitudes by identifying the maximum FFT magnitude within each pass-band for both the model and the filtered observational data. Eight amplitude ratios are computed from the major constituents: M2/S2, M2/N2, S2/N2, (K1+P1)/O1, M2/(K1+P1), M2/O1, S2/(K1+P1), and S2/O1. Each ratio is compared between the model and data, with agreement quantified by the percentage difference: We classify ratio agreement as: Excellent ($<$5\%), Good ($<$10\%), Fair ($<$20\%) or Failed ($>$20\%). Figure~\ref{fig:overall} shows the comparison of amplitude ratios between the Tsoft model and our filtered data, demonstrating close agreement across all ratios.

\begin{figure}
    \centering
    \begin{subfigure}[a]{\columnwidth}
        \centering
        \includegraphics[width=0.4\columnwidth]{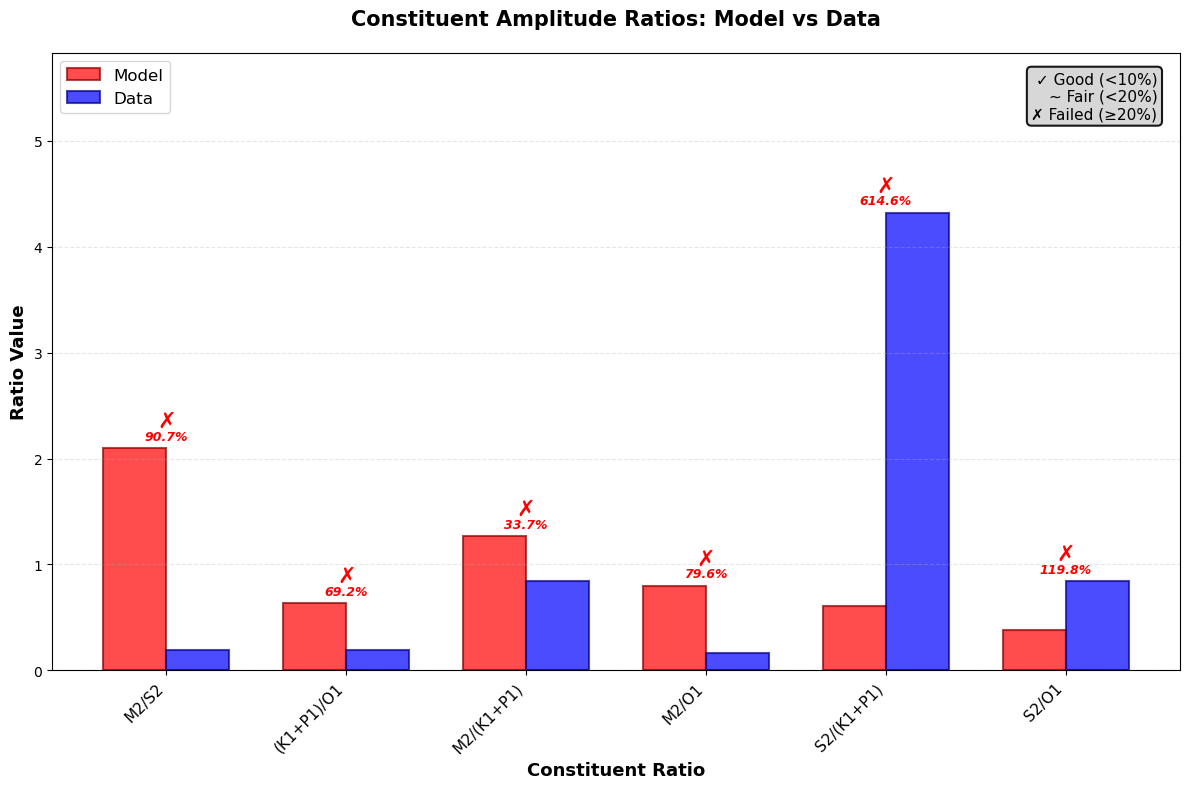}
        \caption{Constituent Amplitude Ratios: Model vs White Noise}
        \label{fig:S6a}
    \end{subfigure}
    \begin{subfigure}[b]{\columnwidth}
        \centering
        \includegraphics[width=0.4\columnwidth]{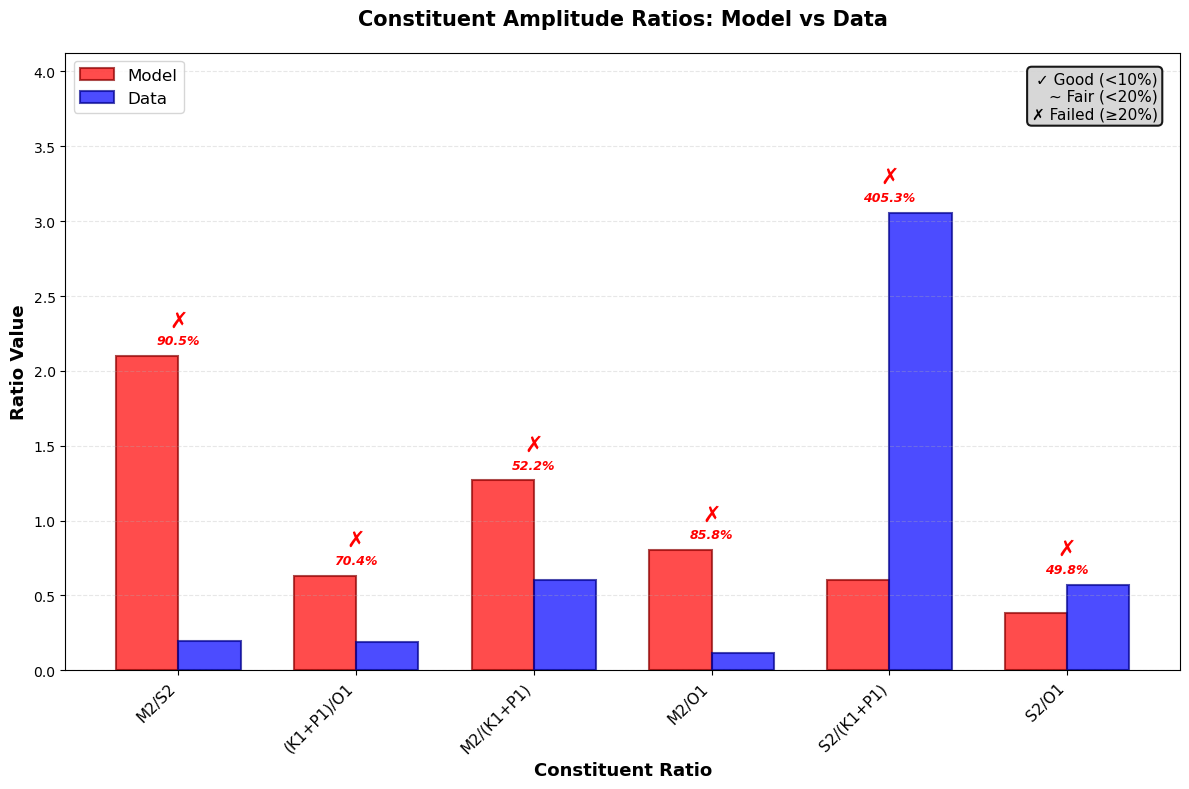}
        \caption{Constituent Amplitude Ratios: Model vs Pink Noise}
        \label{fig:S6b}
    \end{subfigure}
    \begin{subfigure}[c]{\columnwidth}
        \centering
        \includegraphics[width=0.4\columnwidth]{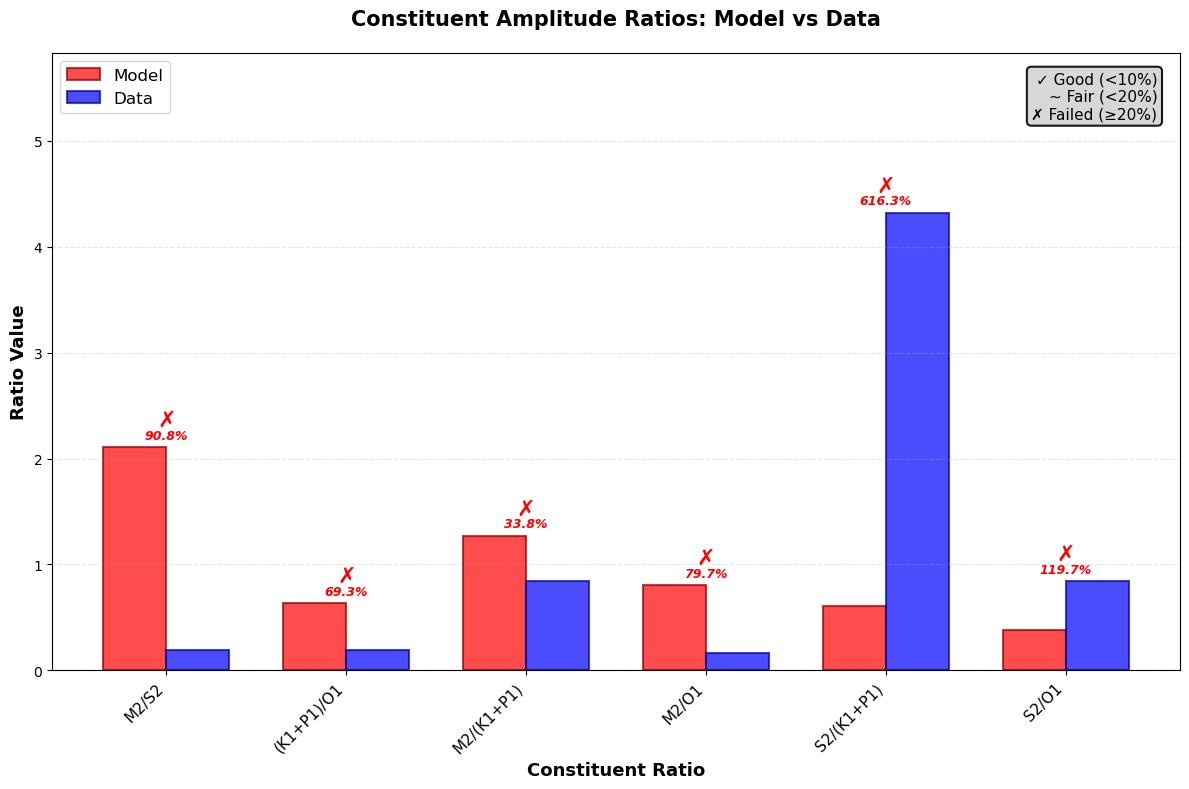}
        \caption{Constituent Amplitude Ratios: Model vs Brown Noise}
        \label{fig:S6c}
    \end{subfigure}
    \label{fig:S6}
    \begin{subfigure}[b]{\columnwidth}
        \centering
        \includegraphics[width=0.4\columnwidth]{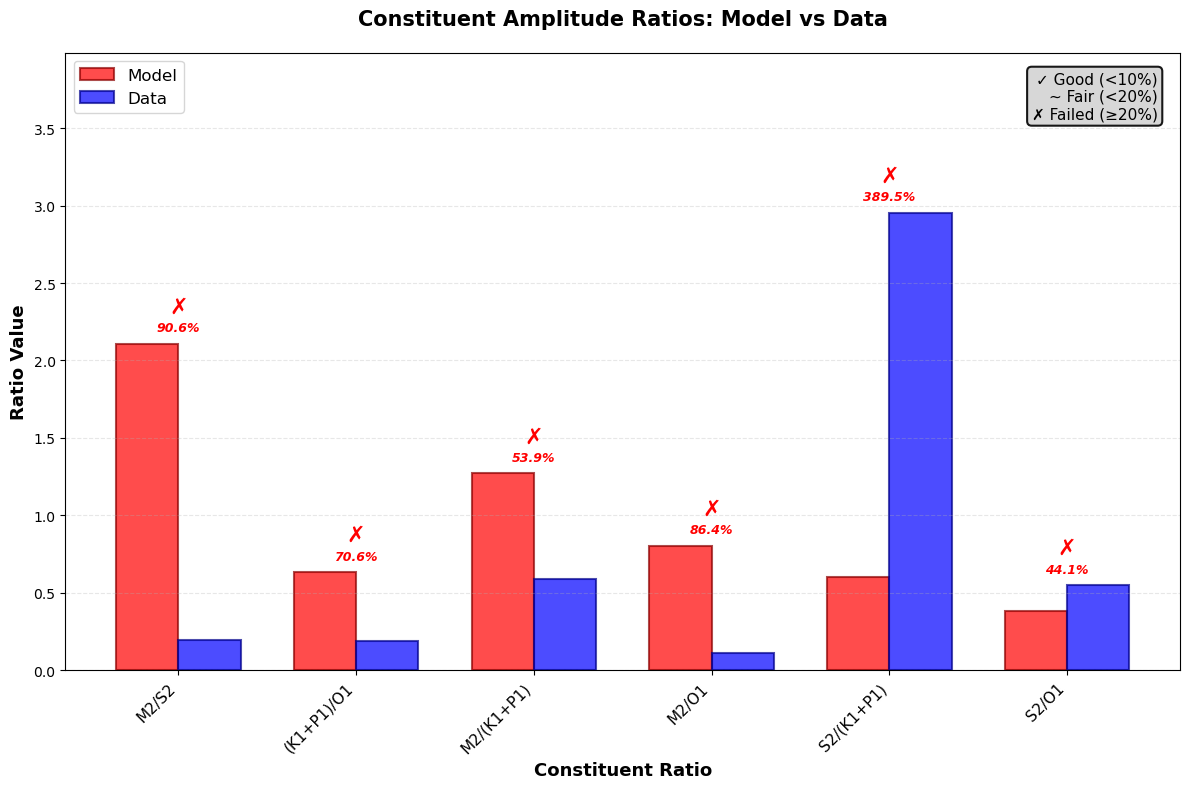}
        \caption{Constituent Amplitude Ratios: Model vs Fitted Noise}
        \label{fig:S6d}
    \end{subfigure}
    \caption{Filter Use Verification: Noise Null Hypothesis Testing. All noise forms tested show clear disagreement between the signal ratios and the filtered noise ratios, indicating that the filtering does not induce artificial correlation but correctly filters out a true signal if present.}
    \label{fig:S6_cont}
\end{figure}

To rigorously verify that the positive ratio matches are not artifacts of the filter design, we performed identical filtering on four types of synthetic noise, each designed to test different potential failure modes. White noise consisting of uncorrelated Gaussian noise with the same standard deviation as our data tests whether random fluctuations could produce false positives (fig.\ref{fig:S6a}). Pink noise ($1/f$, fig.\ref{fig:S6b}) was generated via spectral shaping to produce a power spectral density $S(f) \propto 1/f$, representing flicker noise common in electronic systems. Brown noise ($1/f^2$, fig.\ref{fig:S6c}) was generated to produce $S(f) \propto 1/f^2$, corresponding to random walk behavior. This is particularly important as our Allan deviation analysis indicated our data exhibits potential random walk characteristics at long timescales from gradient. Finally, we constructed a fitted noise curve, where a power-law spectrum was fitted to our 672 hour data set, after which the spectral exponent $\alpha$ was extracted, and synthetic noise was created matching this spectrum. This represents a worst-case scenario where the synthetic noise has identical spectral characteristics to our actual data (fig.\ref{fig:S6d}).

For each noise type, we generated time series with the same length and sampling rate as our observational data, applied the identical filtering procedure, extracted constituent amplitudes, and computed all eight amplitude ratios.

All four null hypothesis tests showed complete failure to reproduce the amplitude ratios observed in our data (figure~\ref{fig:overall}). The mean percentage differences between synthetic noise ratios and model ratios were: white noise (167.93\%), pink noise (125.67\%), brown noise (168.27\%), and fitted noise (122.52\%)---all significantly larger than the 7.37\% achieved with our real data. Critically, the quality classifications were predominantly ``Failed'' for synthetic noise.

We consider these null tests a strong demonstration that the filter does not create spurious amplitude relationships, nor do random coincidental alignment of noise with tidal frequencies explain our positive ratio matches. Furthermore, even noise with spectral characteristics matching our data fails to produce tidal-like amplitude ratios and the excellent ratio agreement in our real data constitutes strong evidence for tidal detection.

\section{C.~Analytical model}

    \begin{figure}
        \centering    
F        \includegraphics[width=0.5\columnwidth]{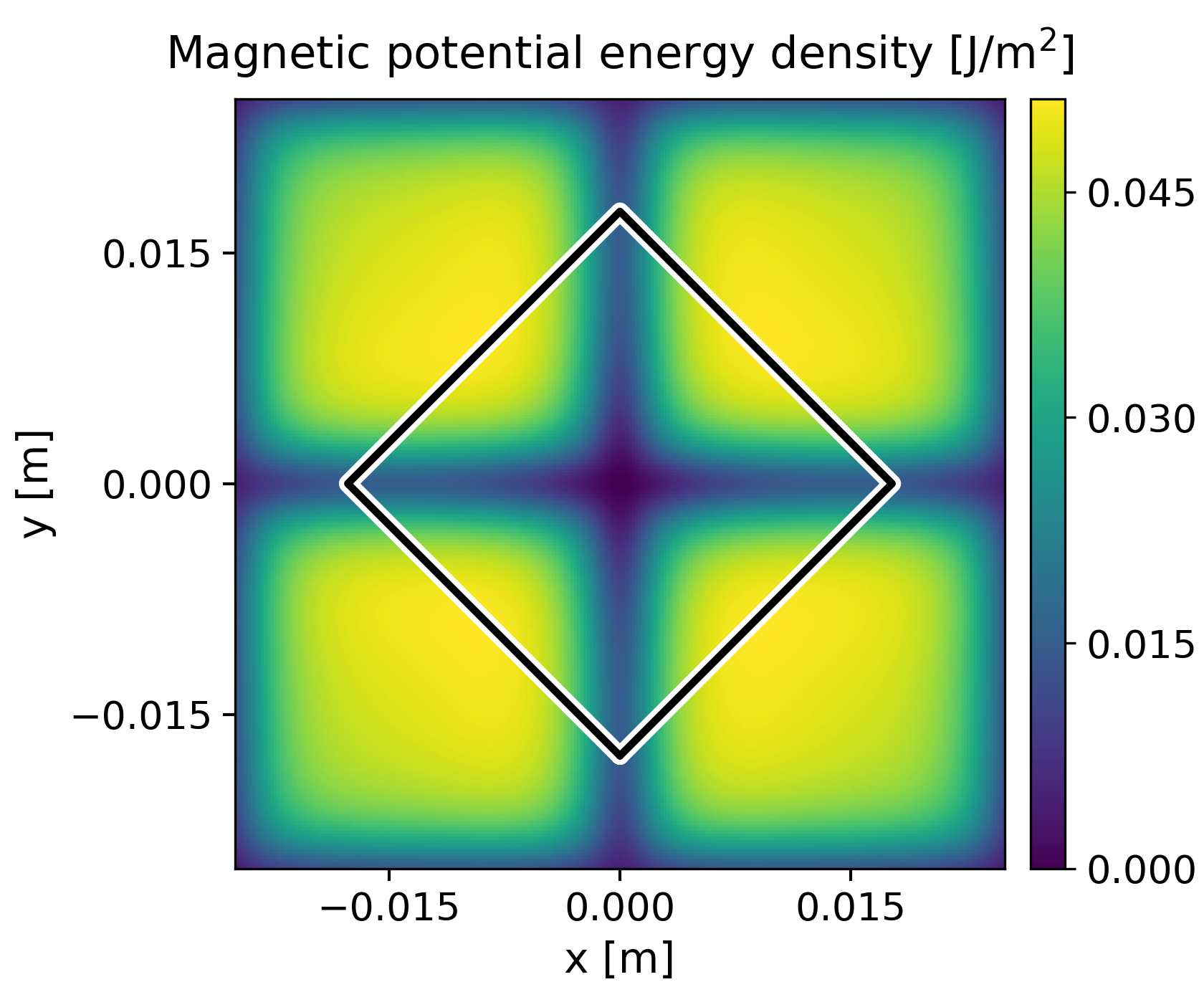}
        \caption{Magnetic potential energy density $U_{\text{mag}}(x,y)$ of the diamagnetic graphite sheet evaluated at the equilibrium levitation height $z_{\text{eq}}$ (base parameters). The overlaid rotated rectangle indicates the graphite  orientation $\phi=\pi/4$.}
        \label{fig:potslice}
    \end{figure}

    \begin{figure*}[!t]
        \hskip0.1cm{\bf A.}\hskip5.2cm{\bf B.}\hskip5.3cm{\bf C.}\\
        \includegraphics[width=0.3\columnwidth]{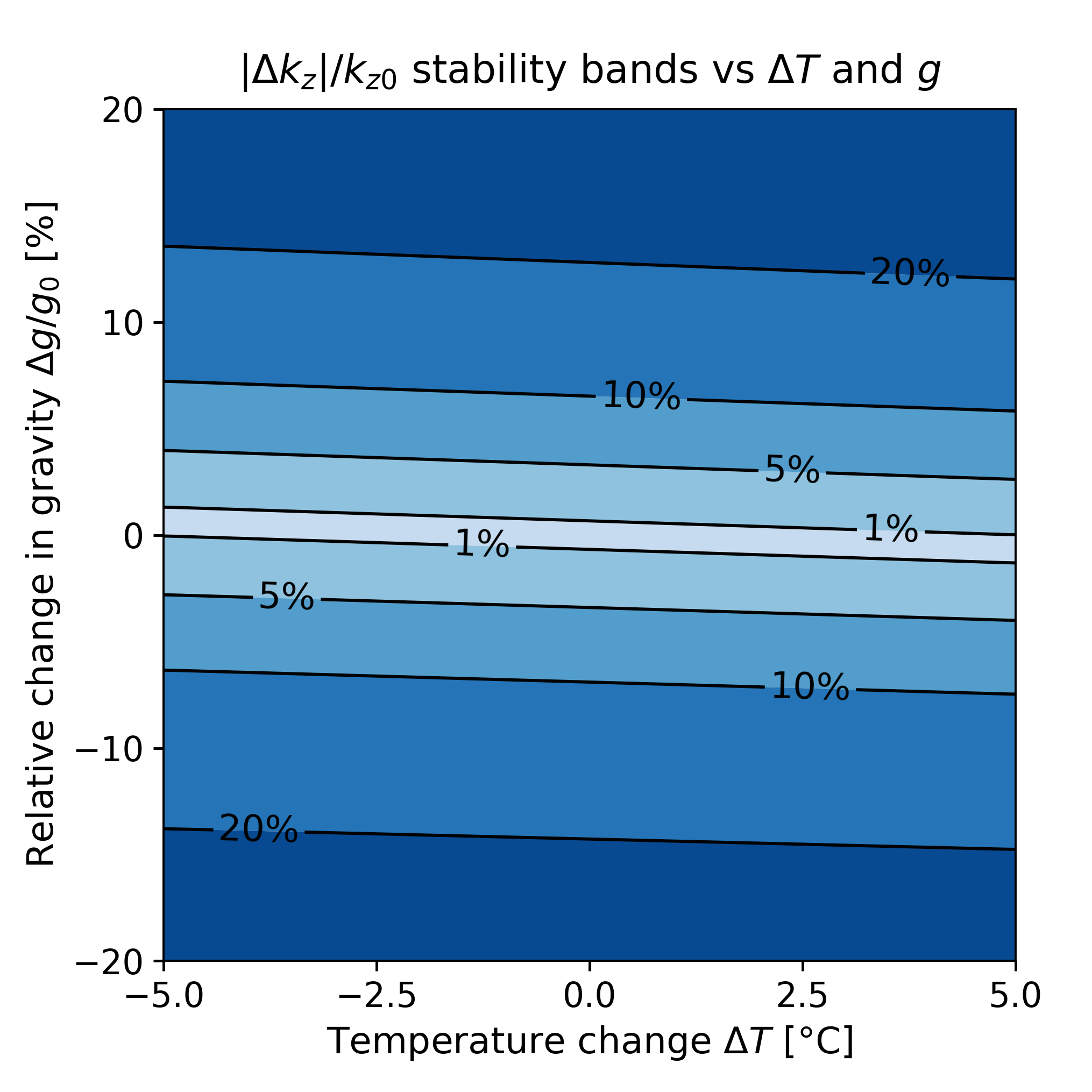}
        \includegraphics[width=0.3\columnwidth]{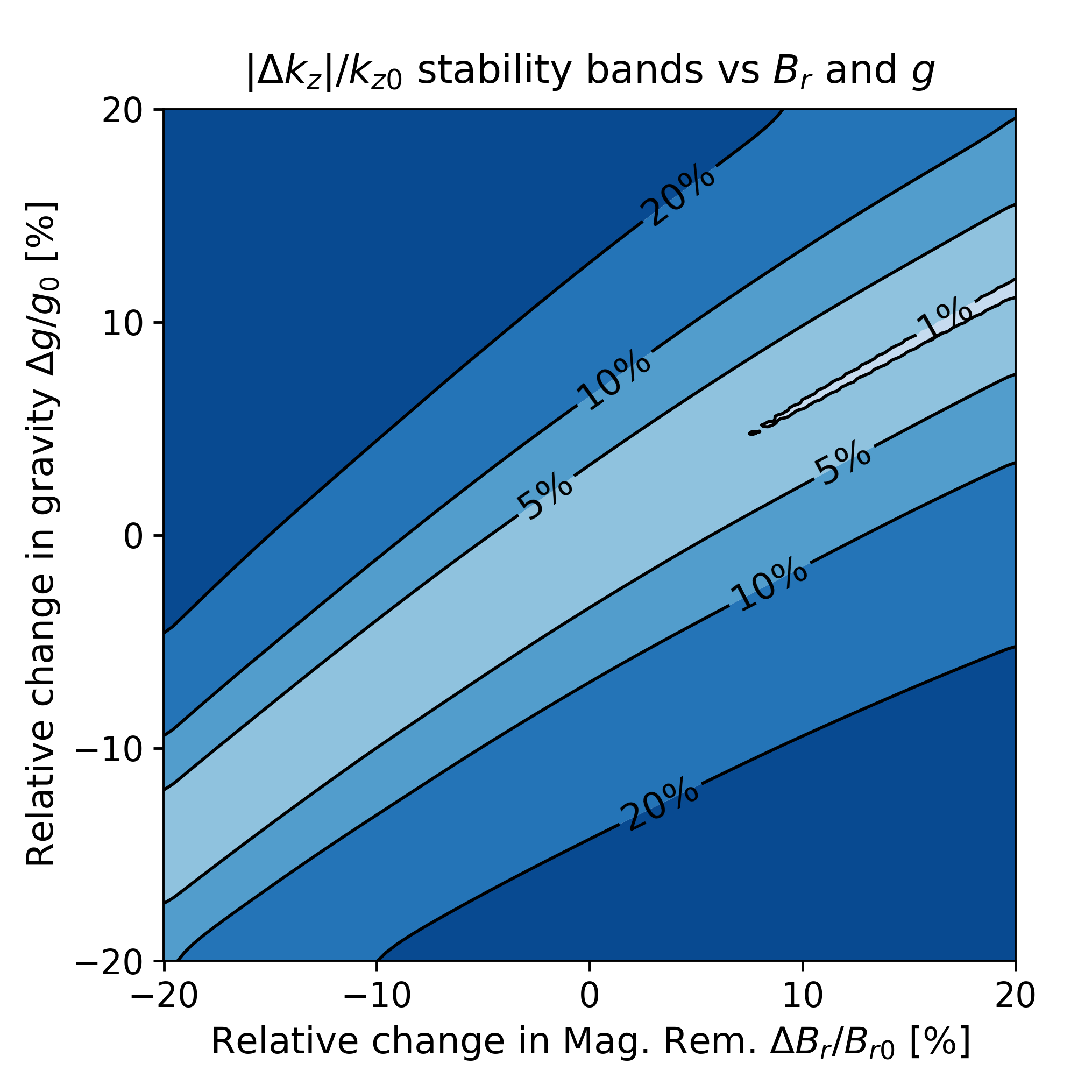}
        \includegraphics[width=0.3\columnwidth]{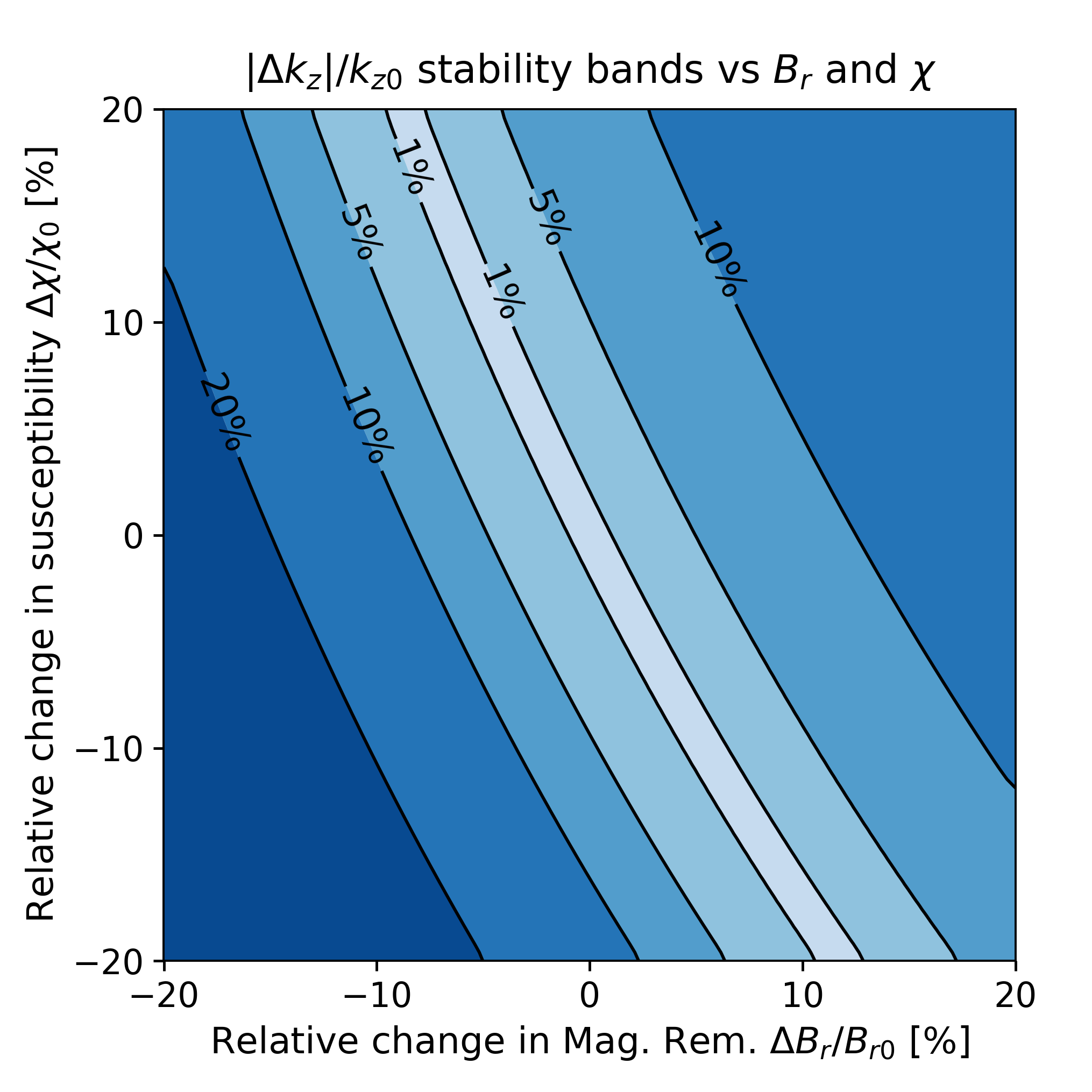} 
        \caption{\textbf{Stability maps from discrete contour bands.} Panels {\bf(A)–(C)} show the relative change of the vertical spring constant $k_z$ across parameter space, plotted as discrete filled contour bands of $100\,|k_z-k_{z0}|/k_{z0}$ (in \%). Here $k_{z0}$ denotes the baseline value at the central operating point of each scan. {\bf (A)} Temperature variation $\Delta T$ versus gravitational acceleration ( $\Delta g/g_0$). {\bf (B)} Remanent field $B_r$ versus $g$. {\bf (C)} $B_r$ versus magnetic susceptibility $\chi$. In all panels, lighter-to-darker shading indicates progressively larger deviations from the baseline stiffness, and the contour boundaries delineate regions of deviation, thereby visualizing the stability of the trap against parameter drifts and identifying the operating region where $k_z$ remains within the chosen tolerance bands.}\label{fig:stability}
\end{figure*}

\noindent
For a cuboidal magnet of side length $2a$, width $2b$ and height $h=2c$ with uniform magnetization $\mathbf{M}=M\,\hat{\mathbf{z}}\,\text{[A/m]}$, where $\hat{\mathbf{z}}$ is the unit vector in the $z$-direction. The magnetic flux density at an arbitrary field point $(x,y,z)$ can be written component-wise as \cite{camacho_2013}:
\begin{align}
B_x(x,y,z) &= \frac{\mu_0 M}{4\pi}\,
\ln\!\left[\frac{F_2(-x,\,y,\,-z)\,F_2(x,\,y,\,z)}{F_2(x,\,y,\,-z)\,F_2(-x,\,y,\,z)}\right],\\[4pt]
B_y(x,y,z) &= \frac{\mu_0 M}{4\pi}\,
\ln\!\left[\frac{F_2(-y,\,x,\,-z)\,F_2(y,\,x,\,z)}{F_2(y,\,x,\,-z)\,F_2(-y,\,x,\,z)}\right],\\[4pt]
B_z(x,y,z) &= -\frac{\mu_0 M}{4\pi}\Big[
F_1(-x,y,z)+F_1(-x,y,-z)\nonumber\\
&\hspace{1.75cm}+F_1(-x,-y,z)+F_1(-x,-y,-z)\nonumber\\
&\hspace{1.75cm}+F_1(x,y,z)+F_1(x,y,-z)\nonumber\\
&\hspace{1.75cm}+F_1(x,-y,z)+F_1(x,-y,-z)\Big],
\end{align}
with auxiliary functions
\begin{align}
F_1(x,y,z) &= \arctan\!\left(\frac{(x+a)(y+b)}{(z+c)\sqrt{(x+a)^2+(y+b)^2+(z+c)^2}}\right),\\[4pt]
F_2(x,y,z) &= \frac{\sqrt{(x+a)^2+(y-b)^2+(z+c)^2}+b-y}{\sqrt{(x+a)^2+(y+b)^2+(z+c)^2}-b-y}.
\end{align}
For a permanent magnet we parameterize $M$ via the magnetic remanence $B_r$ using the standard approximation $\mu_0 M \simeq B_r$. To model the Halbach array, we consider four magnets of equal length and width ($a=b$) in a checkerboard pattern where the four magnets are centered at $(a,a)\,,\,(a,-a)\,,\,(-a,a)$ and $(-a,-a)$ respectively, with magnetization alternating in the positive and negative $z$ direction between $\pm M$. We fix the magnet height to be $z=-c$, such that $z=0$ corresponds to the top of the magnetic array.

\noindent
The levitated sensor is a thin square sheet of graphite of side length $L$ and thickness $\delta$, with a diamagnetic susceptibility vector  $\boldsymbol{\chi}_0=\mathrm{diag}(\chi_{xy},\chi_{xy},\chi_{z}),$  where the $\chi_{xy}$ refer to the magnetic susceptibility parallel to the sheet surface, and $\chi_z$ denotes the magnetic susceptibility normal to the surface of the sheet. The graphite sheet is oriented at an angle of $\pi/4$ in the $x-y$ plane, corresponding to a stable minimum (see Fig \ref{fig:potslice}).

\noindent
The magnetic potential can be calculated by integrating the diamagnetic energy density over the volume of the graphite. For a sufficiently thin sheet, we can approximate its magnetic potential energy by integrating over its surface area
\begin{align}
    U_\text{mag}(x,y,z)=-\dfrac{\delta}{2\mu_0}\int dA&\left[\chi_{xy}\left(B_x^2+B_y^2\right)+\chi_z\,B_z^2\right]
\end{align}
where we have assumed that due to the thinness of the sheet, the magnetic field is approximately the same at the top and the bottom of the sheet. Including a linear gravitational term $mgz$, the total potential energy is
\begin{equation}
U(x,y,z)= m g\,z + U_{\mathrm{mag}}(x,y,z).
\label{eq:Utotal}
\end{equation}
The equilibrium position of the center of the sheet  \((x_{\rm eq},y_{\rm eq},z_{\rm eq})\) can be easily found via the potential
\begin{equation}
\frac{\partial U}{\partial x}=0,\qquad
\frac{\partial U}{\partial y}=0,\qquad
\frac{\partial U}{\partial z}=0,
\label{eq:equilibrium_conditions}
\end{equation}
The effective spring constants are simply the second derivatives of the potential evaluated at equilibrium:
\begin{equation}
k_x=\left.\frac{\partial^2 U}{\partial x^2}\right|_{\rm eq},\quad
k_y=\left.\frac{\partial^2 U}{\partial y^2}\right|_{\rm eq},\quad
k_z=\left.\frac{\partial^2 U}{\partial z^2}\right|_{\rm eq}.
\label{eq:spring_constants}
\end{equation}
The corresponding trap frequencies are
\begin{equation}
\omega_x=\sqrt{\frac{k_x}{m}},\qquad
\omega_y=\sqrt{\frac{k_y}{m}},\qquad
\omega_z=\sqrt{\frac{k_z}{m}},
\label{eq:trap_frequencies}
\end{equation}
and \(f_i=\omega_i/(2\pi)\). From the symmetry of the set-up, $k_x=k_y$.

The potential and its associated effective spring constants depend implicitly on the system parameters. With the dimensions of the graphite sheet and Halbach array fixed, the remaining degrees of freedom are $(B_r,\chi_{xy},\chi_z,g,m)$. 

The remnant magnetization and the diamagnetic susceptibility have further temperature dependence. We also calculate the fractional change in the spring constant as a function of the temperature. The remnant magnetization $B_r$ has been seen to change by around \SI{-0.12}{\%/\celsius} while below \SI{100}{\celsius}, furthermore the diamagnetic susceptibility changes by \SI{-0.15}{\%/\celsius} \cite{magnetochemistry4040052}. 

By modulating the value of these parameters we are able to extract the scaling of the spring constant numerically. While analytical approaches may be possible, the exact form of the equilibrium position which would be extracted from solving \eqref{eq:Utotal}, as well as the exact form of the potential, is non-trivial. As such, a numerical approach was implemented to generate the magnetic potential and associated spring constants. 
\begin{equation}
    U_{mag}\propto\chi\,B_r^2 \label{eq:scaling}
\end{equation}
We emphasize that the scaling in \eqref{eq:scaling} applies to the magnetic potential evaluated at a fixed position (and fixed geometry). The trap stiffnesses \(k_i\), however, are curvatures evaluated at the equilibrium configuration, \(k_i=\left.\partial_i^2 U\right|_{\mathbf{q}=\mathbf{q}_{\rm eq}}\), and the equilibrium itself is parameter-dependent through the condition \(\left.\partial_z U\right|_{z=z_{\rm eq}}=0\) (equivalently \(mg+\left.\partial_z U_{\rm mag}\right|_{z=z_{\rm eq}}=0\)). Varying \(B_r\), \(\chi\), or \(g\) therefore changes both the overall strength of \(U_{\rm mag}\) and the location \(z_{\rm eq}\) at which the spring constant in calculated. In other words, our scans compare different parameter choices at their respective equilibrium heights, rather than comparing different parameter choices at a common, fixed height.

In the numerical implementation we first evaluate \(U_{\rm mag}(z)\) at a reference parameter set by performing the surface integral in \eqref{eq:Utotal} using adaptive two-dimensional quadrature at a sequence of heights \(z\in[z_{\min},z_{\max}]\) (with the sheet orientation fixed to \(\phi=\pi/4\)). These values are then represented by a smooth spline interpolant \(\widetilde U_{\rm mag}(z)\). Parameter scans are generated by exploiting the multiplicative scaling of the magnetic energy with susceptibility and remanence, i.e. 
\begin{equation}
       U_{mag}\to \widetilde{U}_{mag}\equiv\left(\dfrac{\chi'}{\chi}\right)\left(\dfrac{B_r'}{B_r}\right)^2U_{mag} \label{U_scale} ,
\end{equation}
while gravity enters through \(mgz\). For each parameter point we determine the equilibrium height by minimizing \(U(z)=mgz+s\,\widetilde U_{\rm mag}(z)\) in one dimension, using the previous solution as an initial bracket to track the same stable branch. The vertical spring constant is then extracted from the local curvature of the minimized potential, \(k_z\simeq \left.\partial_z^2 U\right|_{z_{\rm eq}}\), computed numerically by a symmetric finite difference about \(z_{\rm eq}\), and the trap frequency follows from \(\omega_z=\sqrt{k_z/m}\) (and \(f_z=\omega_z/2\pi\)). 

\begin{table}[t]
\centering
\caption{Baseline parameters used for the magnetic-potential spline and subsequent scans.}
\label{tab:base_params}
\resizebox{0.8\columnwidth}{!}{%
\begin{tabular}{lll}
\hline
Parameter & Value & Description \\
\hline
$\mu_0$ & $4\pi\times 10^{-7}\ \mathrm{H\,m^{-1}}$ & Vacuum permeability \\
$\chi_{xy}$ & $-8.2\times 10^{-5}$ & Susceptibility ($z$-direction) \\
$\chi_{z}$ & $-5.82\times 10^{-4}$ & Susceptibility ($xy$-direction) \\
$\rho$ & $1442\ \mathrm{kg\,m^{-3}}$ & Graphite density \\
$L$ & $2.5\times 10^{-2}\ \mathrm{m}$ & Graphite side length \\
$\delta$ & $1.0\times 10^{-3}\ \mathrm{m}$ & Graphite thickness \\
$m_0$ & $1.19\times 10^{-3}\ \mathrm{kg}$ & Oscillator mass (Graphite + Mirror) \\
$a$ & $1.25\times 10^{-2}\ \mathrm{m}$ & Magnet half-length \\
$c$ & $6.25\times 10^{-3}\ \mathrm{m}$ & Magnet half-thickness \\
$B_{r0}$ & $1.48\ \mathrm{T}$ & Baseline remanence \\
$M_0$ & $B_{r0}/\mu_0 \approx 1.18\times 10^{6}\ \mathrm{A\,m^{-1}}$ & Baseline magnetization \\
$g_0$ & $9.81\ \mathrm{m\,s^{-2}}$ & Baseline gravitational acceleration \\
\hline
\end{tabular}%
}
\end{table}

\paragraph{Scaling laws}
By curve fitting to the numerically generated data points we are able to extract an approximate scaling laws for the fractional spring constant shown below:

\begin{align}
    \Delta k_z&\approx -0.506  \,\Delta B_r + 0.587 \,(\Delta B_r)^2  -3.597  \,(\Delta B_r)^3 ,\\
    \Delta k_z&\approx -0.308 \,\Delta\chi + 0.368\,(\Delta\chi)^2  -0.205 \,(\Delta\chi)^3 ,\\      
   \Delta k_z&\approx 1.282 \,\Delta g + 0.288 (\,\Delta g)^2 + 0.0217 \,(\Delta g)^3 ,\\   
  \Delta k_z&\approx 0.0014\, \Delta T  -1.58\times 10^{-5}\, (\Delta T)^2 + 5.654\times 10^{-8}\,(\Delta T)^3 ,
\end{align}
where $\Delta A=(A-A_0)/A_0$ for each quantity $A$.

\end{document}